\documentclass[journal,letterpaper,twoside,twocolumn]{IEEEtran}
\usepackage{amsmath,amssymb,graphicx,psfrag,cite}
\usepackage[normalem]{ulem} 
\usepackage[usenames,dvipsnames]{color} 
\usepackage{balance}
\usepackage{mathrsfs}   
\usepackage{paralist}
\usepackage{bbm}

\graphicspath{{figure/}}

\newcommand{\argmax}{\mathop{\rm argmax}\limits}

\newtheorem{theorem}{Theorem}

\usepackage{color}

\newcommand{\rev}[1]{{\color{magenta}#1}}

  \renewcommand{\sout}[1]{}  \renewcommand{\rev}[1]{#1}

\title{Post-FEC BER Benchmarking for Bit-Interleaved Coded Modulation with Probabilistic Shaping} 

\author{Tsuyoshi~Yoshida,~\IEEEmembership{Member,~IEEE,} Alex~Alvarado,~\IEEEmembership{Senior Member,~IEEE,} \\ 
 Magnus~Karlsson,~\IEEEmembership{Fellow,~OSA;~Senior~Member,~IEEE,} and Erik~Agrell,~\IEEEmembership{Fellow,~IEEE} \thanks{Manuscript received October 31, 2019\rev{; r}evised February 17, 2020\rev{; revised April 2, 2020.}}%
	\thanks{This work was presented in part at the 2017 European Conference on Optical Communication \cite{yoshida_ecoc_2017}. It was supported in part by the Swedish Research Council (VR) under grant no. 2017-03702 and ``Massively Parallel and Sliced Optical Network (MAPLE),'' the Commissioned Research of National Institute of Information and Communications Technology (NICT), Japan (project number 20401). The work of A. Alvarado is supported by the Netherlands Organization for Scientific Research (NWO) via the VIDI Grant ICONIC (project number 15685) and has received funding from the European Research Council (ERC) under the European Union's Horizon 2020 research and innovation programme (grant agreement No 757791).}
	\thanks{T.~Yoshida was a visiting researcher of the Fiber Optic Communications Research Center (FORCE), Chalmers University of Technology, SE-41296 Gothenburg, Sweden. He is with Information Technology R\&D Center, Mitsubishi Electric Corporation, Kamakura, Kanagawa, 247-8501 Japan. Currently he also belongs to Graduate School of Engineering, Osaka University, Suita, Osaka, 505-0871 Japan (e-mail: Yoshida.Tsuyoshi@ah.MitsubishiElectric.co.jp).}%
	\thanks{A.~Alvarado is with the Information and Communication Theory Lab, Signal Processing Systems Group, Department of Electrical Engineering, Eindhoven University of Technology, Eindhoven 5600 MB, The Netherlands.}
	\thanks{M.~Karlsson and E.~Agrell are with Fiber Optic Communications Research Center (FORCE), Chalmers University of Technology, SE-41296 Gothenburg, Sweden.}
	\thanks{Copyright (c) 2020 IEEE. Personal use of this material is permitted. However, permission to use this material for any other purposes must be obtained from the IEEE by sending a request to pubs-permissions@ieee.org.}
}%

\begin{document}
\maketitle

\begin{abstract}
Accurate performance benchmarking after forward error correction (FEC) decoding is essential for system design in optical fiber communications. Generalized mutual information (GMI) has been shown to be successful at benchmarking the bit-error rate (BER) after FEC decoding (post-FEC BER) for systems with soft-decision (SD) FEC without probabilistic shaping (PS). However, GMI is not relevant to benchmark post-FEC BER for systems with SD-FEC and PS. For such systems, normalized GMI (NGMI), asymmetric information (ASI), and achievable FEC rate have been proposed instead. They are good at benchmarking post-FEC BER or to give an FEC limit in bit-interleaved coded modulation (BICM) with PS, but their relation has not been clearly explained so far.
In this paper, we define generalized L-values under mismatched decoding, which are connected to the GMI and ASI.
We then show that NGMI, ASI, and achievable FEC rate are theoretically equal under matched decoding but not under mismatched decoding.
We also examine BER before FEC decoding (pre-FEC BER) and ASI over Gaussian and nonlinear fiber-optic channels with approximately matched decoding. ASI always shows better correlation with post-FEC BER than pre-FEC BER for BICM with PS. On the other hand, post-FEC BER can differ at a given ASI when we change the bit mapping, which describes how each bit in a codeword is assigned to a bit tributary.
\end{abstract}

\begin{IEEEkeywords}
	Bit error rate, bitwise decoding, bit-interleaved coded modulation, forward error correction, generalized mutual information, modulation, mutual information, optical fiber communication, probabilistic shaping.
\end{IEEEkeywords}

\section{Introduction}
\label{sec:intro}
In fiber-optic communications, forward error correction (FEC) is widely deployed due to the severe requirements of the residual bit error rate (BER) of the received data, which can be as low as $<10^{-15}$. In research and system evaluations, actual encoding and decoding gives the most reliable performance assessment.  
However, it requires significant efforts to examine very low BERs and the results are specific to the examined FEC code. If one could quantify the performance without actual FEC decoding, it would greatly simplify system development and characterization. Historically, a so-called \emph{FEC limit}, a certain value of the BER before the FEC decoder (hence referred to as pre-FEC BER), has been employed.
It benchmarks the post-FEC (after decoder) performance, to quantify the real available data rates, and to measure margins in deployed systems. 
On the other hand, information-theoretic tools \cite{agrell_2018_ipc} have become increasingly common in fiber-optic communications after the emergence of soft-decision (SD) FEC, which is today mainly used in coherent detection systems with digital signal processing (DSP). For example, transmission techniques 
are often evaluated using achievable information rates (AIRs), which give an indication of the potential data rate assuming ideal FEC performance. 
We should also consider the gap between the AIR and system throughput (sometimes called ``information rate'') and how it is constrained by nonideal FEC performance \cite{cho_2018_ecoc,cho_2019_jlt,vassilieva_2019_ecoc}. 
Thus how information-theoretic metrics like AIR relate to the system throughput in practically relevant cases needs to be understood in detail.

This paper focuses on bit-interleaved coded modulation (BICM) \cite{zehavi_1992_tcom,caire_1998_tit,fabregas_2008,bicmbook} with binary SD FEC, because it has been widely deployed due to its low implementation complexity and design simplicity. We combine BICM with \emph{probabalistic shaping} (PS) with reverse concatenation of FEC and PS which shapes the modulation symbol amplitudes.
This is known as probabilistic amplitude shaping (PAS), and it can approach the Shannon capacity as shown in \cite{bocherer_2015,bocherer_2019}.\footnote{The PS module was called ``distribution matching (DM)'' in the  PAS architecture \cite{bocherer_2015} and generalized to ``PS encoding/decoding'' for layered PS architecture in \cite{bocherer_2019}. There are many PS coding techniques, e.g., \cite{schulte_2016,gultekin_2017,yoshida_ecoc_2018,fehenberger_2019,yoshida_jlt_2019,cho_2019_tcom,goossens_2019_oecc,amari_2019_arxiv}. They are characterized by the (average) rate $k_{\text{ps}}/n_{\text{ps}}$ and average probability mass function (pmf) $\hat{P}_{|X|}$\rev{, where the notations will be explained in Sec.~\ref{sec:system}}.}
Several reports benchmark the post-FEC BER by employing code-independent metrics. For BICM, an AIR is the so-called generalized mutual information (GMI) \cite{kaplan_1993_aeu,bicmbook} under a bitwise reception, and its normalized value, ranging from 0 to 1, has been shown to be a good benchmark of post-FEC BER \cite[Sec.~IV]{alvarado_2015}. In contrast, pre-FEC BER is not good for very low code rates, i.e., pre-FEC BER vs. post-FEC BER curves will differ significantly between low- and high-order modulation formats \cite[Figs.~3(a), 8(a)]{alvarado_2015}. 
A similar metric must be redefined for PAS due to the dependence between bit tributaries and the nonuniform symbol constellations. It has been found that a normalized GMI (NGMI), suitably extended to PAS \cite{cho_ecoc2017,cho_2019_jlt}, or asymmetric information (ASI), defined from bit-weighted log-likelihood ratios
\cite{yoshida_ptl_2017,yoshida_ecoc_2017}, are good benchmark metrics for these systems. These metrics have been used in, e.g., 
\cite{maher_ecoc_2017,chien_ecoc_2018}.
Both NGMI and ASI cover also the uniform (non-PS) BICM cases, which makes these metrics more general than the GMI. 

Three important aspects of these metrics have not yet been fully covered in the literature \cite{alvarado_2015,cho_ecoc2017,yoshida_ptl_2017,yoshida_ecoc_2017,bocherer_2017,cho_2019_jlt,yoshida_ecoc_2018,yoshida_jlt_2019}. 
Firstly, the sometimes subtle relationships between the used rates and metrics should be clarified. 
For example, the relationship between the (normalized) GMI \cite{alvarado_2015,cho_ecoc2017,cho_2019_jlt}, ASI \cite{yoshida_ptl_2017,yoshida_ecoc_2017,yoshida_ecoc_2018,yoshida_jlt_2019}, and achievable FEC rate\cite{bocherer_2019} for PAS has not been sufficiently explained and clarified.  
Recently, NGMI and ASI have been compared for the first time and found to differ \cite{zhang_ofc_2019}. 
Secondly, the lack of studies of the post-FEC performance benchmarking after nonlinear fiber-optic transmission for PAS.
And thirdly, there are yet no investigations of the dependence of post-FEC BER on bit mapping. 
When several bits are mapped onto a modulated symbol, each bit tributary can have different pre-FEC performances. 
For example, the least significant bit has the worst performance in a regular binary reflected Gray-coded pulse amplitude modulated (PAM) signal. 
Thus, the post-FEC BER may differ depending on the bit mapping (how a bit in the FEC codeword is mapped to a bit tributary), even if the other conditions are the same, 
as implied in \cite{hager_2014,bocherer_2015}. More generally, the bit mapping dependence comes from practical nonideal FEC codes, so code-independent 
performance metrics cannot take it into account when the bit mapping is chosen not to change \emph{a priori} probabilities \cite[Remark~5]{bocherer_2019}. Note that density evolution or exit chart analysis, which can take a practical code into account, would be useful in some cases though they are limited only to the specific code used.

This work extends our previous works \cite{yoshida_ptl_2017,yoshida_ecoc_2017}, where we examined the usefulness of ASI under fiber-optic channels under different nonlinear simulation conditions, by addressing the unsolved aspects due to not the main subjects in the literature listed above. The novelties of this paper are  
\begin{itemize}
	\item a generalized definition of L-values, which is connected to channel mismatch, GMI, and ASI, 
	\item a detailed analysis of the relations between performance metrics of NGMI, ASI, and achievable FEC rate, and,
	\item investigation of performance metrics as post-FEC BER benchmark with different bit mappings and various nonlinear transmission cases.
\end{itemize}

We stress that the above metrics relates to post-FEC BER. We do however recognize that the most relevant system output error rates are those after PS decoding in PAS systems, and that the PS decoding tends to increase BER, as shown in \cite{yoshida_ecoc_2018,yoshida_jlt_2019,amari_2019_arxiv} (e.g., 5 to 200 times \cite[Fig.~4]{yoshida_jlt_2019} and 10 to 50 times \cite[Fig.~12]{amari_2019_arxiv}).
The relationship between BERs before and after PS decoding depends on the specific cross layout of FEC and PS codes \cite[Figs.~3]{yoshida_jlt_2019}. Since the BER after PS decoding can be estimated from the post-FEC BER \cite[Sec.~VII]{yoshida_jlt_2019}, this paper, in the interest of generality, focuses only on performance metrics before FEC decoding that estimates the post-FEC BER. 

The rest of the paper is organized as follows. Relevant metrics are summarized in Sec.~\ref{sec:theo}, where the equivalence between NGMI and ASI is discussed in detail. 
Bit mapping dependence and performance benchmarking over the nonlinear fiber-optic channel are shown in Secs.~\ref{sec:bl_mapg} and \ref{sec:fo_ch}, resp. Sec.~\ref{sec:cncl} concludes the paper.

\section{Theory for relevant rates and metrics}
\label{sec:theo}
In this section, after definition of the system model and newly defined generalized L-values (Sec.~\ref{sec:system}), we summarize achievable rates (Sec.~\ref{sec:air}), auxiliary channels (Sec.~\ref{sec:aux_ch}), decoding metrics (Sec.~\ref{sec:dec_met}), and relevant performance metrics before and after FEC decoding (Sec.~\ref{sec:metric}). Then the relation of information rate and FEC code rate is explained (Sec.~\ref{sec:rates}), which is useful for understanding the metrics. Finally we discuss the equivalence of the NGMI, ASI, and achievable FEC rate (Sec.~\ref{sec:eq_metric}).\footnote{\rev{Our analysis starts from the well-known measure of GMI in Sec.~\ref{sec:air}, which is subsequently expanded into auxiliary channels and decoding metrics. An alternative approach would be to start from the general \cite[Th.~2]{bocherer_2019} and instantiate it for specific scenarios.}}

\subsection{System model and generalized L-values}
\label{sec:system}
\begin{figure}[t]
	\begin{center}
		\setlength{\unitlength}{.6mm} %
		\scriptsize
		\includegraphics[scale=0.45]{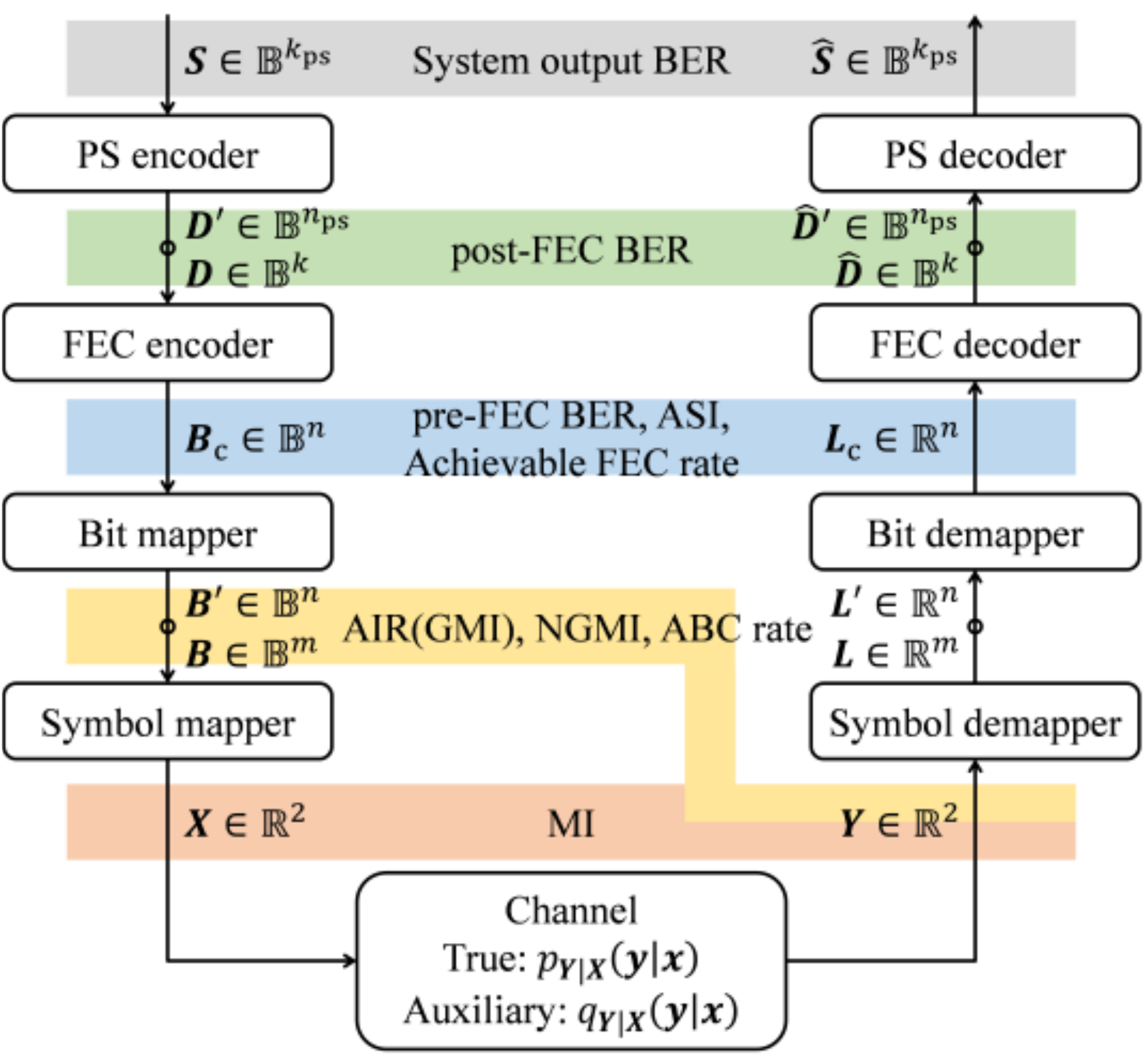}\\
		\vspace{-0.3cm}
		\caption{System model of BICM with PS. Here we show the notations of the signals with the number of the dimensions and key performance metrics that will be discussed in this paper.}
		\label{fig:system}
	\end{center}
	\vspace{-0.3cm}
\end{figure}

We consider the system model shown in Fig.~\ref{fig:system}. At the transmitter, the source bits $\boldsymbol{S}\in \mathbb{B}^{k_{\text{ps}}}$ are assumed to be uniformly distributed and independent. 
The source bits are converted to the shaped bits $\boldsymbol{D}' \in \mathbb{B}^{n_{\text{ps}}}$ by a PS encoder based on
PAS\footnote{In this work we focus on PAS, though reverse concatenation PS is more general, e.g., it can comprise different coding/shaping schemes like multi-level coding. PAS shapes absolute amplitudes only in each dimension and does not shape sign bits\cite{bocherer_2015}.}, where $\mathbb{B} = \{ 0,1 \}$ and $(n_{\text{ps}}, k_{\text{ps}})$ are the numbers of PS encoder output and input bits per PS codeword. Each circle on the arrows in Fig.~\ref{fig:system} denotes block length conversion, e.g., $\mathbb{B}^{n_{\text{ps}}}$ to $\mathbb{B}^{k}$. 
A systematic FEC encoding is applied to the payload bits $\boldsymbol{D} \in \mathbb{B}^{k}$ for PAS, whereby the encoded bits $\boldsymbol{B}_{\text{c}} \in \mathbb{B}^{n}$ are generated, where $(n, k)$ are the FEC codeword length and number of information bits per FEC codeword. The bits $\boldsymbol{B}_{\text{c}}$ are then mapped to bit tributaries in $\boldsymbol{B}' \in \mathbb{B}^{m\cdot (n/m)}$, where $m$ denotes the number of bit tributaries conveyed by each two-dimensional symbol, and $n/m$ is assumed to be an integer for simplicity. 
Next the bits $\boldsymbol{B} = B_1 \ldots B_m \in \mathbb{B}^{m}$ are mapped into a two-dimensional symbol $\boldsymbol{X} \in \mathcal{X} \in \mathbb{R}^{2}$, where $\mathcal{X}$ and $\mathbb{R}$ denote the set of two-dimensional symbols and real numbers, resp. 
In an FEC codeword, a shaping set $\mathcal{S}$ is $\in \mathcal{X}^{N}$, where $|\mathcal{S}|=2^{k_{\text{ps}}k/n_{\text{ps}}}$ and $N=n/m$.
After being transmitted over the channel, the received symbol $\boldsymbol{Y} \in \mathbb{R}^{2}$ 
is demapped to bitwise \emph{a posteriori} L-values, also known as log-likelihood ratios $\boldsymbol{L} =L_1 \ldots L_m \in \mathbb{R}^m$. The discrete-time, memoryless channel law $p_{\boldsymbol{Y}|\boldsymbol{X}}(\boldsymbol{y}|\boldsymbol{x})$ will in the following be called the \emph{true channel,} although the memory in fiber-optic channel is neglected.

The L-values $\boldsymbol{L}' \in \mathbb{R}^{m\cdot (n/m)}$ are bit-demapped to $\boldsymbol{L}_{\text{c}} \in \mathbb{R}^{n}$, which are decoded to $\hat{\boldsymbol{D}} \in \mathbb{B}^{k}$ to recover the shaped bits. The PS decoding is performed on $\hat{\boldsymbol{D}}' \in \mathbb{B}^{n_{\text{ps}}}$ when PAS is employed. Finally the source bits are recovered as $\hat{\boldsymbol{S}} \in \mathbb{B}^{k_{\text{ps}}}$.
For uniform signaling $\boldsymbol{D}=\boldsymbol{S}$ and $\hat{\boldsymbol{S}} = \hat{\boldsymbol{D}}$.

Under mismatched decoding in a bitwise receiver, the mismatched L-values are called \emph{generalized} L-values \cite[Th.~4.20]{bicmbook}. In this work, we explicitly define the generalized L-values applicable to systems with PS for the first time. This is useful to describe and understand the scaling of L-values, channel assumed in the demapper, decoding metric, GMI in \eqref{eq:GMIs}, and ASI in \eqref{eq:ASI}.
The generalized L-values are calculated per bit tributary $i$ from the received symbol $\boldsymbol{Y}$. The generalized \emph{a posteriori} L-value $\hat{L}_i^{\text{po}}(\boldsymbol{Y})$ is defined as 
\begin{IEEEeqnarray}{rCL}
	\label{eq:Lvalue}
	\hat{L}_i^{\text{po}}(\boldsymbol{y}) & \triangleq & L_i^{\text{pr}} + s \hat{L}_i^{\text{ex}}(\boldsymbol{y}) . 
\end{IEEEeqnarray}
The scaling parameter $s$ is useful to correct the L-values, as done in \cite[Ch.~7]{bicmbook}, \cite{alvarado_2016,nguyen_2011,szczecinski_2012}.
Here, $L_i^{\text{pr}}$ and $\hat{L}_i^{\text{ex}}$ are \emph{a priori} and generalized \emph{extrinsic} L-values, i.e., 
\begin{IEEEeqnarray}{rCL}
	\label{eq:Lpr}
	L_i^{\text{pr}} &\triangleq& \ln \frac{P_{B_i}(0)}{P_{B_i}(1)} , \\
	\label{eq:Lex}
	\hat{L}_i^{\text{ex}}(\boldsymbol{y}) &\triangleq& \ln \frac{q_{\boldsymbol{Y} \mid B_i}(\boldsymbol{y} \! \mid \! 0)}{q_{\boldsymbol{Y} \mid B_i}(\boldsymbol{y} \! \mid \! 1)} ,
\end{IEEEeqnarray}
where $B_i$ and $P_{B_i}(b_{i})$ denote bits for bit tributary index $i$ and the probability of bits $B_i$ being $b_{i}\in\mathbb{B}$, and $\boldsymbol{b}=b_1 \ldots b_m$. The notations $q_{B_i,\boldsymbol{Y}}(b_{i},\boldsymbol{y})$ and $q_{\boldsymbol{Y} | B_i}(\boldsymbol{y} | b_{i})$ are the joint and conditional probability density funcitons (pdf:s) assumed in the demapper, resp., of $\boldsymbol{Y}$ and $B_i$. 
We note here that in general the bitwise ``auxiliary channel'' $q_{\boldsymbol{Y} | B_i}(\boldsymbol{y} | b_{i})$ in \eqref{eq:Lex} is different from the auxiliary channel $q_{\boldsymbol{Y} | \boldsymbol{B}}(\boldsymbol{y} | \boldsymbol{b})$, the bitwise true channel $p_{\boldsymbol{Y} | B_i}(\boldsymbol{y} | b_{i})$, and the true channel $p_{\boldsymbol{Y} | \boldsymbol{B}}(\boldsymbol{y} | \boldsymbol{b})$ (see Fig.~\ref{fig:system}).
Scaling of $\hat{L}_i^{\text{ex}}$ by the scaling parameter $s$ in \eqref{eq:Lvalue} is equivalent to adjusting the bitwise auxiliary channel to $q_{\boldsymbol{Y} | B_i}^{(\text{s})}(\boldsymbol{y} | b_{i}) \propto q_{\boldsymbol{Y} | B_i}(\boldsymbol{y} | b_{i})^s$. Then, the generalized \emph{a posteriori} L-values in \eqref{eq:Lvalue} are rewritten with the scaled and assumed \emph{a posteriori} probability $q_{B_i \mid \boldsymbol{Y}}^{(\text{s})}(b_i \! \mid \! \boldsymbol{y})$, i.e.,
\begin{IEEEeqnarray}{rCL}
	\label{eq:Lpo}
	\hat{L}_i^{\text{po}}(\boldsymbol{y}) &=& \ln \frac{q_{B_i \mid \boldsymbol{Y}}^{(\text{s})}(0 \! \mid \! \boldsymbol{y})}{q_{B_i \mid \boldsymbol{Y}}^{(\text{s})}(1 \! \mid \! \boldsymbol{y})} , \\
	q_{B_i \mid \boldsymbol{Y}}^{(\text{s})}(b_i \! \mid \! \boldsymbol{y}) & = &  \frac{P_{B_i}(b_i)q_{\boldsymbol{Y}\mid B_i}^{(\text{s})}(\boldsymbol{y}\!\mid\! b_i)}{q_{\boldsymbol{Y}}(\boldsymbol{y})} ,
\end{IEEEeqnarray}
where $q_{\boldsymbol{Y}}(\boldsymbol{y})$ denotes the assumed pdf of the received symbol.
The L-values $\hat{L}_1^{\text{po}}(\boldsymbol{Y}) \ldots \hat{L}_m^{\text{po}}(\boldsymbol{Y})$ are used in place of $\boldsymbol{L}=L_1\ldots L_m$, which is bit demapped and fed to the FEC decoder.\footnote{In a deployable demapper and an FEC decoder, the demapper output L-value $L_i({\boldsymbol{Y}})$ is quantized to, e.g., three to six bits. Then another scaling before the quantization works to change the quantization step $\Delta l$. This will be discussed in Sec\rev{s}.~\rev{\ref{sec:dec_met}} and \ref{sec:eq_metric}\rev{,} and Appendix~C.}

In a BICM scheme, the bit mapper/demapper in Fig.~\ref{fig:system} is usually described as bit interleaver/de-interleaver. The bit interleaver shuffles the bit positions \cite[Sec.~2.7]{bicmbook} and determines the bit tributary $i$ for each bit. An important motivation for the shuffling is to scatter burst errors caused by the channel, or by an inner decoder (if there is one). In this work we mainly focus on how a bit in the FEC codeword is assigned to a bit tributary in the symbol, which is independent of burst errors.\footnote{We simply set the bit mapping period to the FEC codeword length $n$ in order to maintain the pmf $P_{|X|}$. This is same as the interleave length in \cite[Remark~5]{bocherer_2019}. In a practical receiver we can choose a different interleave length if we keep $P_{|X|}$ unchanged.}

\subsection{Achievable rates}
\label{sec:air}
While the mutual information (MI) $I(\boldsymbol{X} ; \boldsymbol{Y})$ is the highest AIR for coded modulation, GMI is an AIR for receivers using mismatched decoding \cite[Eq.~(20)]{kaplan_1993_aeu}. 
It is defined for any memoryless channel as
\begin{IEEEeqnarray}{rCL}
	\label{eq:GMI}
	\!\!\!\! \text{GMI} &\triangleq& I_{q,s_{\text{o}}}^{\text{gmi}} (\boldsymbol{B}; \boldsymbol{Y}), \\
	\label{eq:GMI_so}
	\!\!\!\! s_{\text{o}} & \triangleq & \argmax_{s \ge 0} I_{q,s}^{\text{gmi}} (\boldsymbol{B}; \boldsymbol{Y}),\\
	\label{eq:GMIs}
	\!\!\!\! I_{q,s}^{\text{gmi}} (\boldsymbol{B}; \boldsymbol{Y}) &\triangleq& \mathbb{E}_{\boldsymbol{B},\boldsymbol{Y}} \!\! \left[ \log_2 \frac{q_{\boldsymbol{Y} \mid \boldsymbol{B}}(\boldsymbol{Y} \! \mid \! \boldsymbol{B})^s}{\sum_{\boldsymbol{b}\in\mathbb{B}^m} P_{\boldsymbol{B}}(\boldsymbol{b}) q_{\boldsymbol{Y} \mid \boldsymbol{B}}(\boldsymbol{Y} \! \mid \! \boldsymbol{b})^s} \right],
\end{IEEEeqnarray}
where the label $\boldsymbol{B}$ and the symbol $\boldsymbol{X}$ are a one-to-one mapping, $\mathbb{E}_{\boldsymbol{B},\boldsymbol{Y}} \left[ \cdot \right]$ denotes the expectation under the joint probability $p_{\boldsymbol{B},\boldsymbol{Y}}(\boldsymbol{b},\boldsymbol{y})$, and $q_{\boldsymbol{Y} | \boldsymbol{B}}(\boldsymbol{y} | \boldsymbol{b})$ denotes an auxiliary channel.\footnote{Note that in \eqref{eq:GMIs} $q_{\boldsymbol{Y}|\boldsymbol{B}}(\boldsymbol{y}|\boldsymbol{b})$ is the auxiliary channel and not the decoding metric $\mathbbm{q}(\boldsymbol{b},\boldsymbol{y})$ in \cite[Eq.~(4.35)]{bicmbook}. See details in Secs.~\ref{sec:aux_ch} and \ref{sec:dec_met}.} 
The scaling parameter $s$ in \eqref{eq:GMIs} is discussed in Sec.~\ref{sec:system}. Under general matched decoding (i.e., $q_{\boldsymbol{Y} | \boldsymbol{B}}(\boldsymbol{y} | \boldsymbol{b}) = p_{\boldsymbol{Y} | \boldsymbol{B}}(\boldsymbol{y} | \boldsymbol{b})$), the optimum scaling parameter $s_{\text{o}}$ in \eqref{eq:GMI_so} is $1$.

The decoding in a bitwise receiver is called bit metric decoding (BMD).
When PAS is added to BICM, an AIR is the BMD rate \cite[Eq.~(51)]{bocherer_2017}
\begin{IEEEeqnarray}{rCL}
	\label{eq:RBMDY}
	R_{\text{bmd}} &\triangleq& \left[\mathbb{H}(\boldsymbol{B}) -\sum_{i=1}^{m} \mathbb{H}(B_i \! \mid \! \boldsymbol{Y}) \right]^+.  
\end{IEEEeqnarray}
In \eqref{eq:RBMDY} $\mathbb{H}(\cdot)$, $\mathbb{H}(\cdot | \cdot)$, and $[ \cdot ]^{+}$ denote entropy, conditional entropy, and $\text{max} \{ \cdot , 0 \}$, resp.
While the operation $[ \cdot ]^{+}$ prevents $R_{\text{bmd}}$ from being negative, the net difference without $[ \cdot ]^{+}$,
\begin{IEEEeqnarray}{rCL}
	\label{eq:ent_diff}
	\Delta_{\mathbb{H}} \triangleq \mathbb{H}(\boldsymbol{B})-\sum_{i=1}^{m}\mathbb{H}(B_i \! \mid \! \boldsymbol{Y}) 
\end{IEEEeqnarray}
will also be useful in the analysis.

In \cite[Eqs.~(12), (13), Appendix]{cho_2019_jlt}, there is no scaling parameter in the definition of GMI \eqref{eq:GMI}, so $s_{\text{o}}=1$ is implicitly assumed.  In that case,
\begin{IEEEeqnarray}{rCL}
	\label{eq:gmi_deltaH}
	\text{GMI} = \Delta_{\mathbb{H}}
\end{IEEEeqnarray}
as derived in \cite{cho_2019_jlt}. 

In \cite[Eq.~(104)]{bocherer_2019}, an AIR is defined with shaping set size $|\mathcal{S}|$, number of symbols $N$, and uncertainty $\mathbb{U}$, i.e.,
\begin{IEEEeqnarray}{C}
	\label{eq:AIR_unc}
	\text{AIR} = \left[ \frac{\log_2 |\mathcal{S}|}{N} - \mathbb{U} \right]^{+} .
\end{IEEEeqnarray}
Substituting $|\mathcal{S}|=2^{k_{\text{ps}}k/n_{\text{ps}}}$ and $N=n/m$ (see Sec.~\ref{sec:system}) into \eqref{eq:AIR_unc}, we obtain
\begin{IEEEeqnarray}{C}
	\text{AIR} = \left[ \frac{k_{\text{ps}}k}{n_{\text{ps}}n}m - \mathbb{U} \right]^{+} .
\end{IEEEeqnarray}
The quantity $\log_2 | \mathcal{S} | / N$ gives an upper bound on the information rate. \rev{The} \rev{u}ncertainty $\mathbb{U}$ \cite[Sec.~V-B]{bocherer_2019} will be explained in Sec.~\ref{sec:metric}. The description  \eqref{eq:AIR_unc} includes the rate loss in PS, which will be discussed in Sec.~\ref{sec:rates}.

\subsection{Auxiliary channels}
\label{sec:aux_ch}
The auxiliary channel $q_{\boldsymbol{Y}|\boldsymbol{B}}(\boldsymbol{y}|\boldsymbol{b})$ in \eqref{eq:GMIs} is described with an \emph{a posteriori} pdf assumed in the demapper $q_{\boldsymbol{B} | \boldsymbol{Y}}(\boldsymbol{b} | \boldsymbol{y})$ as
\begin{IEEEeqnarray}{rCL}
	\label{eq:auxch_symbol}
	q_{\boldsymbol{Y} \mid \boldsymbol{B}}(\boldsymbol{y} \! \mid \! \boldsymbol{b}) & = & \frac{q_{\boldsymbol{Y}}(\boldsymbol{y})}{P_{\boldsymbol{B}}(\boldsymbol{b})} q_{\boldsymbol{B} \mid \boldsymbol{Y}}(\boldsymbol{b} \! \mid \! \boldsymbol{y}).
\end{IEEEeqnarray}
In a bitwise receiver, $q_{\boldsymbol{B} | \boldsymbol{Y}}(\boldsymbol{b} | \boldsymbol{y})$ is the product of bitwise \emph{a posteriori} probabilities assumed in the demapper, i.e.,
\begin{IEEEeqnarray}{rCL}
	\label{eq:apoPr}
	q_{\boldsymbol{B} \mid \boldsymbol{Y}}(\boldsymbol{b} \!\mid \!\boldsymbol{y}) & = & \prod_{i=1}^m q_{B_i \mid \boldsymbol{Y}}(b_i \!\mid \!\boldsymbol{y}). 
\end{IEEEeqnarray}
In \eqref{eq:apoPr}, $q_{B_i | \boldsymbol{Y}}(b_i | \boldsymbol{y})$ is used to define the generalized \emph{a posteriori} L-value $\hat{L}_i^{\text{po}} (\boldsymbol{y})$ in \eqref{eq:Lpo}. 
From \eqref{eq:auxch_symbol} and \eqref{eq:apoPr},\footnote{This expression is the same as  \cite[Eq.~(22)]{buchali_2016}.}
\begin{IEEEeqnarray}{rCL}
	\label{eq:aux_ch}
	q_{\boldsymbol{Y} \mid \boldsymbol{B}}(\boldsymbol{y} \! \mid \! \boldsymbol{b}) & = & \frac{q_{\boldsymbol{Y}}(\boldsymbol{y})}{P_{\boldsymbol{B}}(\boldsymbol{b})} \prod_{i=1}^m q_{B_i \mid \boldsymbol{Y}}(b_i \! \mid \! \boldsymbol{y}), \\ 
	\label{eq:aux_ch2}
	&=&\frac{\prod_{i=1}^m P_{B_i}(b_i)}{P_{\boldsymbol{B}}(\boldsymbol{b})}\prod_{i=1}^m q_{\boldsymbol{Y} \mid B_i}(\boldsymbol{y} \! \mid \! b_{i}) .
\end{IEEEeqnarray}
Considering the definitions of generalized L-values in \eqref{eq:Lvalue}--\eqref{eq:Lex}, i.e.,
\begin{IEEEeqnarray}{rCL}
	\hat{L}_i^{\text{po}}(\boldsymbol{y}) & = & \ln \frac{P_{B_i}(0)}{P_{B_i}(1)} + \ln \frac{q_{\boldsymbol{Y}\mid B_i}^{\rev{(\text{s})}}(\boldsymbol{y}\!\mid\! 0)}{q_{\boldsymbol{Y}\mid B_i}^{\rev{(\text{s})}}(\boldsymbol{y}\!\mid\! 1)} ,
\end{IEEEeqnarray}
\eqref{eq:aux_ch2} can be expressed as
\begin{IEEEeqnarray}{rCL}
	\label{eq:aux_chs}
	q_{\boldsymbol{Y} \mid \boldsymbol{B}}(\boldsymbol{y} \! \mid \! \boldsymbol{b})^{s} & = & \frac{\prod_{i=1}^m P_{B_i}(b_i)}{P_{\boldsymbol{B}}(\boldsymbol{b})}\prod_{i=1}^m q_{\boldsymbol{Y} \mid B_i}^{\rev{(\text{s})}}(\boldsymbol{y} \! \mid \! b_i) , 
\end{IEEEeqnarray}
where scaling is applied only to the bitwise auxiliary channel, and \emph{a priori} probabilities are not scaled.\footnote{Similar discussion is found in \cite[Sec.~6.1]{bocherer_2017}. We keep further investigation for a future work.}

The bitwise auxiliary channel $q_{\boldsymbol{Y} \mid B_i}(\boldsymbol{y} \! \mid \! b)$ is typically given by \cite[Eq.~(11)]{buchali_2016}
\begin{IEEEeqnarray}{rCL}
	\label{eq:bw_auxch}
	q_{\boldsymbol{Y} \mid B_i}(\boldsymbol{y} \! \mid \! b) &=& \frac{1}{P_{B_i}(b)} \sum_{\boldsymbol{b} \in \mathbb{B}^m: b_i = b} P_{\boldsymbol{B}}(\boldsymbol{b}) q_{\boldsymbol{Y} \mid \boldsymbol{B}}^{\text{awgn}} (\boldsymbol{y} \! \mid \! \boldsymbol{b}) .
\end{IEEEeqnarray}
The pdf $q_{\boldsymbol{Y} | \boldsymbol{B}}^{\text{awgn}}(\boldsymbol{y} | \boldsymbol{b})$ denotes the assumed Gaussian channel 
\begin{IEEEeqnarray}{rCL}
	q_{\boldsymbol{Y} \mid \boldsymbol{B}}^{\text{awgn}} (\boldsymbol{y} \! \mid \! \boldsymbol{b}) & = & \frac{1}{\sqrt{2\pi \hat{\sigma}^2}}\exp \left\{ - \frac{(\boldsymbol{y} - \boldsymbol{x}({\boldsymbol{b}}))^2}{2 \hat{\sigma}^2} \right\} ,
\end{IEEEeqnarray}
where $\boldsymbol{x}({\boldsymbol{b}})$ denotes channel input $\boldsymbol{x}$ with label $\boldsymbol{b}$. The assumed standard deviation of the noise  $\hat{\sigma}$ is given by
\begin{IEEEeqnarray}{rCL}
	\hat{\sigma} & = & \sqrt{ \sum_{\boldsymbol{b}\in \mathbb{B}^m} P_{\boldsymbol{X}}(\boldsymbol{x}) |\boldsymbol{x}|^2 / \widehat{\text{SNR}} } ,
\end{IEEEeqnarray}
where $\widehat{\text{SNR}}$ denotes assumed channel SNR. Under the mismatched Gaussian channel, $s_{\text{o}}$ in \eqref{eq:GMI_so} is given by SNRs in true and assumed channels, i.e.,
\begin{IEEEeqnarray}{rCL}
	\label{eq:Gauss_so}
	s_{\text{o}} & = & \text{SNR} / \widehat{\text{SNR}}.
\end{IEEEeqnarray}

\subsection{Decoding metrics}
\label{sec:dec_met}

The arbitrary decoding metric is denoted by $\mathbbm{q}(\cdot,\cdot)$ to distinguish it from the other parameters. As shown in \cite[Eq.~(57)]{bocherer_2019}, an ideal FEC decoder will recover the codeword $\hat{\boldsymbol{D}}$ so that
\begin{IEEEeqnarray}{rCL}
	\label{eq:decmet}
	\hat{\boldsymbol{D}}=\argmax_{\boldsymbol{D} \in \mathbb{B}^k} \mathbbm{q}(\boldsymbol{B}_{\text{c}}, \boldsymbol{Y}_{\text{c}} ) ,
\end{IEEEeqnarray}
where $\boldsymbol{Y}_{\text{c}}$ denotes received symbols $\boldsymbol{Y}^{n/m}$ from the FEC encoded bits $\boldsymbol{B}_{\text{c}}$ from the information bits $\boldsymbol{D}$ in an FEC codeword.
The decoding in a bitwise receiver is called BMD.
In a memolyless bitwise receiver,\footnote{In this work, the FEC-encoded bits $\boldsymbol{B}_{\text{c}}$ (length $n$) in a codeword are assumed to be converted to $n/m$ symbols $\boldsymbol{X}^{n/m}$ for simplicity. In a general BICM system, $\boldsymbol{B}_{\text{c}}$ can be dispersed to $n$ symbols and $\boldsymbol{X}^n$ is constructed by FEC encoded bits from multiple codewords.}
\begin{IEEEeqnarray}{rCL}
	\label{eq:decmet_dmc}
	\hat{\boldsymbol{D}} 
	& = & \argmax_{\boldsymbol{D} \in \mathbb{B}^k} \prod_{j=1}^{n/m} \mathbbm{q}_{\text{bmd}}^{\rev{s_{\rev{\text{d}}}}}(\boldsymbol{b}[j],\boldsymbol{y}[j]) , 
\end{IEEEeqnarray}
where $\mathbbm{q}_{\text{bmd}}(\cdot,\cdot)$ denotes the bitwise decoding metric for given transmitted bits $\boldsymbol{b}[j]$ and the received symbol $\boldsymbol{y}[j]$ for the $j$-th symbol in an FEC codeword.
\rev{The optimization of the scaling parameter $s_{\rev{\text{d}}}$ in \eqref{eq:unc_star} works to minimize the uncertainty, and helps achieving a tighter bound.}
The bitwise decoding metric is given by \cite[Eq.~(79)]{bocherer_2019}
\begin{IEEEeqnarray}{rCL}
	\label{eq:qbmd_lin}
	\mathbbm{q}_{\text{bmd}}(\boldsymbol{b},\boldsymbol{y}) & = & \prod_{i=1}^m q_{B_i \mid \boldsymbol{Y}}(b_i \! \mid \! \boldsymbol{y}) ,
\end{IEEEeqnarray}
which coincides the right-hand side of \eqref{eq:apoPr}.
Alternatively, log-domain BMD is employed in practice\footnote{\rev{The linear and log-domain BMDs (\eqref{eq:qbmd_lin} and \eqref{eq:qbmd_log}) are related via an order-preserving transformation \cite[Sec.~8.2]{bocherer_2018}.}\sout{In general, \eqref{eq:decmet_dmc} and \eqref{eq:qbmd_log} are not equivalent.}}, i.e.,
\begin{IEEEeqnarray}{C}
	\label{eq:qbmd_log}
	\hat{\boldsymbol{D}} = \argmax_{\boldsymbol{D} \in \mathbb{B}^k}  \sum_{j=1}^{n/m} \rev{s_{\rev{\text{d}}}}\mathbbm{q}_{\text{bmd,log}}(\boldsymbol{b}[j],\boldsymbol{y}[j]) ,  \\
	\label{eq:decmet_Lval}
	\mathbbm{q}_{\text{bmd,log}}(\boldsymbol{b},\boldsymbol{y}) = \sum_{i=1}^m (-1)^{b_i} l_i(\boldsymbol{y}) , 
\end{IEEEeqnarray}
where $\mathbbm{q}_{\text{bmd,log}}$ denotes log-domain bitwise decoding metric, and $b_i[j]$ and $l_i(\boldsymbol{y}[j])$ are the bit and the decoder input L-value from the $i$-th bit tributary in the $j$-th symbol, resp. \rev{In \eqref{eq:qbmd_log}, the scaling parameter $s_{\text{d}}$ does not influence $\hat{D}$. On the other hand, $s_{\text{d}}$ influences $\hat{D}$ with quantization or truncation of L-values before FEC decoding, which is usual in practical circuitry.}\footnote{\rev{The scaling at the demapper output and that at the FEC decoder input are equivalent. In this work, all such scaling are done at the demapper. See footnote~4, Sec.~\ref{sec:eq_metric}, and Appendix~C about the scaling at the output of the demapper.}}

\subsection{Performance metrics}
\label{sec:metric}
A common performance target in deployable systems is that the post-FEC BER,
\begin{IEEEeqnarray}{rCL}
	\label{eq:postBER}
	\text{BER}_{\text{post}} &\triangleq& \sum_{b\in \mathbb{B}} P_{D,\hat{D}} (b,1-b) ,
\end{IEEEeqnarray}
must be very small, i.e., $\text{BER}_{\text{post}}<10^{-15}$. To evaluate if the obtained performance corresponds to error-free operation, so-called \emph{FEC limits} have been utilized. This concept relies on the assumption that there exists a one-to-one mapping between the post-FEC BER and certain metrics before FEC decoding.
If true it would enable accurate performance estimations in experiments or simulations without implementing FEC.
The most common metric has been the BER before FEC decoding, pre-FEC BER, defined as
\begin{IEEEeqnarray}{rCL}
	\label{eq:preBER}
	\text{BER}_{\text{pre}} &\triangleq& \sum_{b\in \mathbb{B}} P_{B,\text{sign}(L)}(b,1-b) ,
\end{IEEEeqnarray}
where $L \in \mathbb{R}$ (without the index $i$) denotes ``mixed'' L-value in an FEC codeword. 
The variable $L$ therefore denotes an element of the L-value vector before FEC decoding, $\boldsymbol{L}_{\text{c}} \in \mathbb{R}^{n}$.
Historically, this metric was used for hard-decision (HD) FEC, and it is accurate for memoryless HD systems. However, its accuracy is not enough for SD-FEC, particularly at lower code rates \cite[Fig.~3(a), 8(a)]{alvarado_2015}, \cite[Fig.~3]{alvarado_ofc_2015}.
Instead, $\text{GMI}/m$ was introduced and worked well for uniform signaling with SD-FEC.\footnote{If there are burst errors and the FEC codeword length is limited, interleaving over time is required to disperse the errors. If not, the post-FEC BER will be degraded compared with the expectation. More details are given in \cite[Sec.~II-B]{alvarado_2018_jlt}.}
Its value range is limited to [0,1] for uniform signaling.
To extend this to the PAS case, normalization of the BMD rate $R_{\text{bmd}}$ by its maximum value $\mathbb{H}(\boldsymbol{B})$
was introduced as the normalized AIR \cite[Eq.~(7)]{yoshida_ptl_2017}
\begin{IEEEeqnarray}{rCL}
	\label{eq:RBMD_HB}
	\text{Normalized AIR} \triangleq R_{\text{bmd}} / \mathbb{H}(\boldsymbol{B}).
\end{IEEEeqnarray}
The normalized AIR in \eqref{eq:RBMD_HB} does not work for PAS, as shown in \cite{yoshida_ptl_2017}.

Instead of the simple normalization in \eqref{eq:RBMD_HB}, a useful metric is
\begin{IEEEeqnarray}{rCL}
	\label{eq:NGMIps}
	\text{NGMI} & \triangleq & 1-(\mathbb{H}(\boldsymbol{B}) - \text{GMI})/m , \\
	\label{eq:NGMIdef}
	&=& 1 - \left( \mathbb{H}(\boldsymbol{B}) - I_{q,s_{\text{o}}}^{\text{gmi}} (\boldsymbol{B}; \boldsymbol{Y}) \right) /m .
\end{IEEEeqnarray}
The NGMI was first studied in \cite{alvarado_ofc_2015} for uniform signals and later extended to PS in \cite{cho_ecoc2017}, \cite[Eq.~(15)]{cho_2019_jlt}.\footnote{Here \eqref{eq:ent_diff} and \eqref{eq:gmi_deltaH} are applied to \eqref{eq:NGMIps}. In the regime $\Delta_{\mathbb{H}} \ge 0$, $\text{NGMI}=1-\left( \mathbb{H}(\boldsymbol{B})-R_{\text{bmd}} \right) /m$.}  In \eqref{eq:NGMIdef}, we generalize the concept further by including the optimization over $s$, which is important for bitwise mismatched decoding.
The NGMI in \eqref{eq:NGMIdef} with $s_{\text{o}}=1$ can be shown to be the same as the \emph{achievable binary code rate} (ABC rate) defined in \cite[Example~4.1]{bocherer_2017}
\begin{IEEEeqnarray}{rCL}
	\label{eq:ABCR}
	\text{NGMI} = \text{ABC rate} &\triangleq& 1 - \frac{1}{m} \sum_{i=1}^m \mathbb{H}(B_i \! \mid \! \boldsymbol{Y}) .
\end{IEEEeqnarray}
In other words, the NGMI in \eqref{eq:NGMIdef} coincides with the ABC rate when the bitwise auxiliary channel is matched to the bitwise true channel ($q_{\boldsymbol{Y} | B_i}(\boldsymbol{y} |  b)=p_{\boldsymbol{Y} | B_i}(\boldsymbol{y} |  b)$, and thus, $s_{\text{o}}=1$). The ABC rate is therefore a special case of the NGMI, and thus, from now on we will not discuss the ABC rate any further.

The NGMI gives the maximum FEC code rate for error-free operation with an ideal FEC in the PAS scheme, as explained in \cite[Sec.~III]{yoshida_ptl_2017},\cite[Sec.~IV]{cho_2019_jlt}, and also in Sec.~\ref{sec:rates} in this paper.

Independently of the NGMI, the \emph{asymmetric information (ASI)} was introduced in \cite{yoshida_ptl_2017} as
\begin{IEEEeqnarray}{rCL}
	\label{eq:ASI}
	\text{ASI} &\triangleq& 1 - \mathbbm{h}(L_{\text{a}} \! \mid \! |L_{\text{a}}|) = 1 + \mathbbm{h}(|L_\text{a}|) - \mathbbm{h}(L_{\text{a}}),
\end{IEEEeqnarray}
where $\mathbbm{h}(\cdot | \cdot)$ and $\mathbbm{h}(\cdot)$ denote conditional differential entropy and differential entropy, resp.\footnote{ASI can be computed based on the quantized L-values, which is relevant in deployable systems. In such cases, the differential entropy $\mathbbm{h}(\cdot)$ in \eqref{eq:ASI} is replaced by the entropy $\mathbb{H}(\cdot)$.} The \emph{asymmetric L-value} $L_{\text{a}}$ and its pdf are given by
\begin{IEEEeqnarray}{rCL}
	\label{eq:La}
	L_{\text{a}} &\triangleq& (-1)^{B} L , 
\end{IEEEeqnarray}
where $B$ is the transmitted bit, and
\begin{IEEEeqnarray}{rCL}
	\label{eq:PLa}
	p_{L_{\text{a}}}(l) &=& \sum_{b\in \mathbb{B}} P_B(b) p_{L \mid B}((-1)^b \cdot l \! \mid \! b) .
\end{IEEEeqnarray}

\begin{figure}[t]
	\begin{center}
		\setlength{\unitlength}{.6mm} %
		\scriptsize
		\includegraphics[scale=0.45]{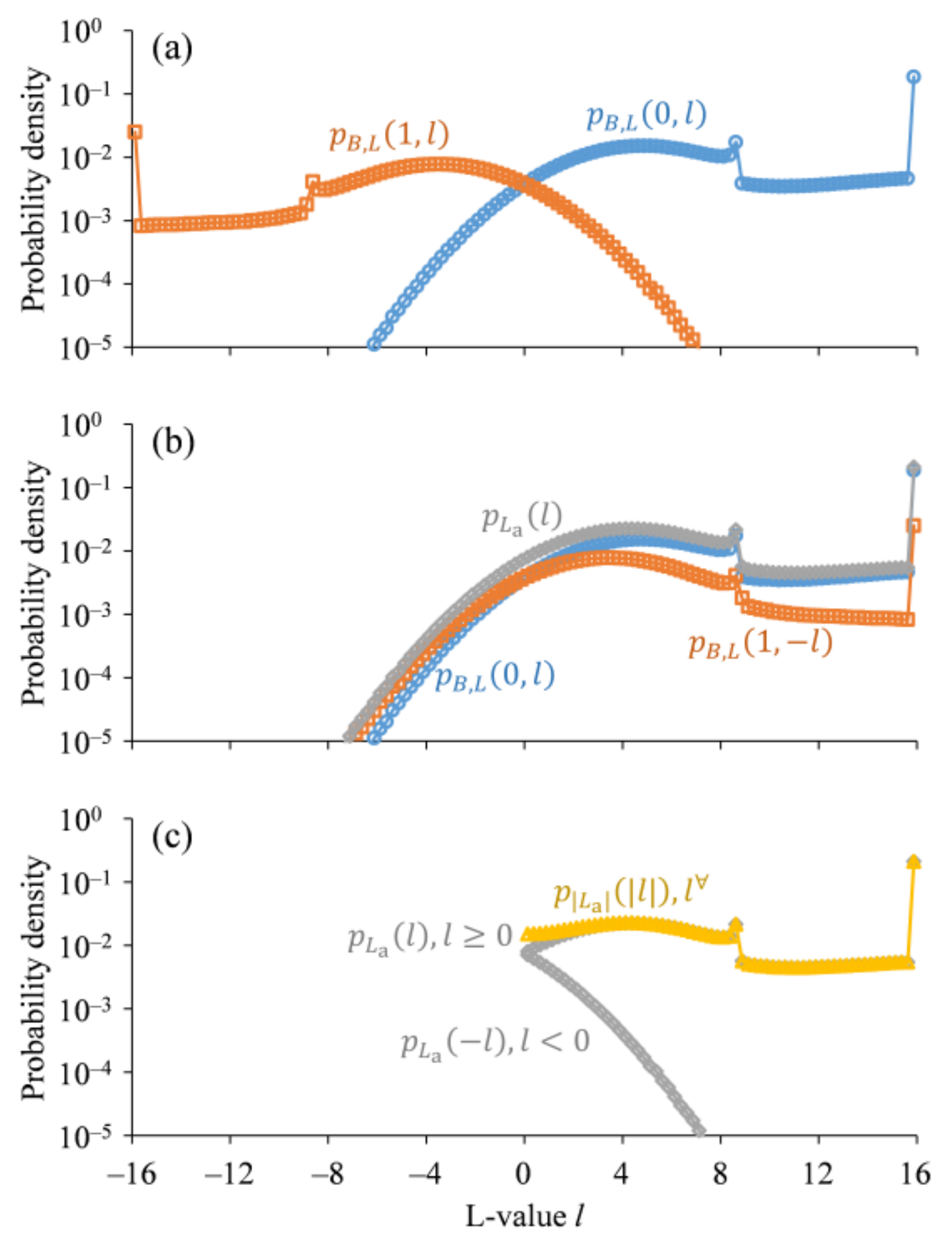}
		\vspace{-0.3cm}
 		\caption{An example of pdfs in the case of PAS-64-QAM (system (i) in Tab.~\ref{tab:sim_8PAM}) at a SNR of 9 dB: (a) $p_{B,L}(b,l)$, (b) $p_{L_{\text{a}}}(l)$, and (c) $p_{|L_{\text{a}}|}(l)$. The high peaks are due to clipping the pdf by a finite range histogram. The effect of the clipping at large $L_{\text{a}}$ in (b) is negligible for calculating the ASI under a reasonable scaling of the L-values.}
		\label{fig:pdf_example}
	\end{center}
	\vspace{-0.3cm}
\end{figure}

Fig.~\ref{fig:pdf_example} shows an example of the pdfs for PAS 64-ary quadrature amplitude modulation (64-QAM), a shaping scheme that will be explained in Sec.~\ref{sec:bl_mapg}. Fig.~\ref{fig:pdf_example}(a) shows the joint distribution $p_{B,L}(b,l)$, Fig.~\ref{fig:pdf_example}(b) the pdfs of the ``symmetrized'' L-values $p_{B,L}(0,l)$, $p_{B,L}(1,-l)$, and $p_{L_{\text{a}}}(l)$ \cite{ivanov16}, and Fig.~\ref{fig:pdf_example}(c) the pdfs of $p_{|L_{\text{a}}|}(|l|)$, $p_{L_{\text{a}}}(l)\,(l\ge 0)$, and $p_{L_{\text{a}}}(-l)\,(l<0)$. The pdf of $p_{L_{\text{a}}}(-l)\,(l<0)$ corresponds to the probability density with hard-decision bit errors.
We showed in \cite{yoshida_ptl_2017,yoshida_ecoc_2017} that the ASI benchmarks the post-FEC BER well for both 
uniform and shaped signaling.

In \cite[Eqs.~(62), (74)]{bocherer_2019}, the achievable FEC rate $R_{\text{fec}}^{*}$ and uncertainty $\mathbb{U}$ for a memoryless channel are defined as
\begin{IEEEeqnarray}{rCL}
	\label{eq:afr}
	\!\!\!\!\!\!\!\!\!\!\!\! R_{\text{fec}}^{*}(\mathbbm{q},\boldsymbol{x}^{\frac{n}{m}},\boldsymbol{y}^{\frac{n}{m}}) & \triangleq & \left[ 1- \frac{\mathbb{U}(\mathbbm{q}, \boldsymbol{x}^{\frac{n}{m}}, \boldsymbol{y}^{\frac{n}{m}})}{m} \right]^{+} , \\
	\label{eq:unc}
	\!\!\!\!\!\!\!\!\!\!\!\! \mathbb{U}(\mathbbm{q},\boldsymbol{x}^{\frac{n}{m}},\boldsymbol{y}^{\frac{n}{m}})  & = &  -\frac{1}{n} \sum_{j=1}^{n/m} \log_2 \frac{\mathbbm{q}(\boldsymbol{x} [j], \boldsymbol{y} [j])}{\sum_{\boldsymbol{v} \in \mathcal{X}} \mathbbm{q}(\boldsymbol{v}, \boldsymbol{y} [j])} , 
\end{IEEEeqnarray}
where the $j$-th transmitted and received symbols in an FEC codeword (length $n$) are $\boldsymbol{x} [j]$ and $\boldsymbol{y} [j]$.
The uncertainty is minimized with a scaling parameter $s_{\rev{\text{d}}}$ as
\begin{IEEEeqnarray}{rCL}
	\label{eq:unc_star}
	\mathbb{U}^{*}(\mathbbm{q},\boldsymbol{x}^{\frac{n}{m}},\boldsymbol{y}^{\frac{n}{m}}) & \triangleq & \min_{s_{\rev{\text{d}}} \rev{\ge} 0} \mathbb{U}(\mathbbm{q}^{s_{\rev{\text{d}}}},\boldsymbol{x}^{\frac{n}{m}},\boldsymbol{y}^{\frac{n}{m}}) , 
\end{IEEEeqnarray}
where $s_{\rev{\text{d}}}$ corresponds to the scaling parameter $s$ for GMI in \eqref{eq:GMI}.  
The uncertainty in a bitwise receiver with the decoding metric $\mathbbm{q}_{\text{bmd}}$ in \eqref{eq:qbmd_lin} is \cite[Eq.~(80)]{bocherer_2019}
\begin{IEEEeqnarray}{rCL}
	\label{eq:unc_b}
	\mathbb{U}(\mathbbm{q}_{\text{bmd}},\boldsymbol{B},\boldsymbol{Y}) & \rev{=} & \sum_{i=1}^m \mathbb{H} (B_i \! \mid \! \boldsymbol{Y}) .
\end{IEEEeqnarray}
The uncertainty with the log-domain bitwise decoding metric $\mathbbm{q}_{\text{bmd,log}}$ in \eqref{eq:decmet_Lval} is expressed as \rev{\cite[Eq.~(86)]{bocherer_2019}}
\begin{IEEEeqnarray}{L}
	\!\!\!\!\!\!\!\! \mathbb{U}(\rev{s_{\text{d}}}\mathbbm{q}_{\text{bmd,log}},\boldsymbol{B}_{\text{c}},\boldsymbol{L}_{\text{c}}) \nonumber \\
	\label{eq:unc_l}
	\!\!\!\!\!\!\!\! = \sum_{i=1}^m \mathbb{E}_{B_i,\hat{L}_i^{\text{po}}} \! \left[ \log_2 (1+\exp \left( -(-1)^{B_i} s_{\rev{\text{d}}} \hat{L}_i^{\text{po}}(\boldsymbol{Y})) \right) \right] \rev{,}
\end{IEEEeqnarray}
\rev{which is minimized by \cite[Eqs.~(87), (100)]{bocherer_2019}}
\begin{IEEEeqnarray}{rCL}
	\label{eq:uncL_star}
	\rev{\mathbb{U}^{*}(\mathbbm{q}_{\text{bmd,log}},\boldsymbol{B}_{\text{c}},\boldsymbol{L}_{\text{c}})} & \rev{=} & \rev{\min_{s_{\text{d}} \rev{\ge} 0} \mathbb{U}(s_{\text{d}}\mathbbm{q}_{\text{bmd,log}},\boldsymbol{B}_{\text{c}},\boldsymbol{L}_{\text{c}}) .}
\end{IEEEeqnarray}

\subsection{Information rate and code rate with rate loss}
\label{sec:rates}
The system throughput (without any overhead) is characterized not only by the AIR, but also by the rate loss due to nonideal FEC\footnote{\label{ft:codegap} The difference between NGMI or ASI and the FEC code rate $R_{\text{c}}$ is called coding gap in \cite[Sec.~IV]{cho_2019_jlt}. It was considered also in \cite{vassilieva_2019_ecoc}, \cite[Tab.~II]{koike_jlt_2017}.} and
PS coding.\footnote{Overheads for framing and pilot signals should be also be accounted for in practice.} 
The information rate $R$, which is the AIR minus FEC and PS coding rate losses, can be written as
\begin{IEEEeqnarray}{rCL}
	\label{eq:R_nu_rc}
	R &\triangleq& \mathbb{H}(\boldsymbol{B}) - R_{\text{loss}} - (1-R_{\text{c}}) m ,
\end{IEEEeqnarray}
where $R_{\text{c}}=k/n$ is the FEC code rate, and $R_{\text{loss}}$ denotes the PS coding rate loss \cite[Sec.~V-B]{bocherer_2015}, \cite[Eq.~(4)]{fehenberger_2019}.
The obtained AIR from the shaping set size and uncertainty in \eqref{eq:AIR_unc} includes the rate loss in PS. At a sufficiently large QAM symbol length $N$ in a PS codeword, $\log_2 |\mathcal{S}|/N$ approaches $\mathbb{H}(\boldsymbol{B})$ with a negligible rate loss. See details in \cite[Sec.~II-D, Sec.~III-C, Th.~1, Sec.~VI-D, Sec.~VII-G]{bocherer_2019}. 
Note that here we define it per two dimensions, as also done in \cite[Eq.~(3)]{yoshida_jlt_2019}.
Example values of $\mathbb{H}(\boldsymbol{B})$, and $R_{\text{loss}}$ can be found in Sec.~\ref{sec:sim_gc}, Tab.~\ref{tab:sim_8PAM}.
The AIR given by the BMD rate \eqref{eq:RBMDY} is also corrected with the PS coding rate loss to \cite[Eq.~(15)]{fehenberger_2019} 
\begin{IEEEeqnarray}{rCL}
	R_{\text{bmd}}' & = & \Delta_{\mathbb{H}} - R_{\text{loss}} \nonumber \\
	&=&
	\label{eq:RBMDL_MDpr}
	\mathbb{H}(\boldsymbol{B}) - \sum_{i=1}^{m} \mathbb{H}(B_i \! \mid \! \boldsymbol{Y}) - R_{\text{loss}}
\end{IEEEeqnarray}
in the regime where $\Delta_{\mathbb{H}} \ge 0$.
The information rate $R$ in \eqref{eq:R_nu_rc} must satisfy $R\leq R_{\text{bmd}}'$, where $R_{\text{bmd}}'$ is given by \eqref{eq:RBMDL_MDpr}. Hence, the FEC code rate satisfies
\begin{IEEEeqnarray}{rCL}
	\label{eq:Rc_bound}
	R_{\text{c}} &\le& 1 - \left( \mathbb{H}(\boldsymbol{B}) - R_{\text{loss}} - R_{\text{bmd}}' \right) / m . 
\end{IEEEeqnarray}
The right-hand side of \eqref{eq:Rc_bound} is equal to the NGMI in \eqref{eq:ABCR}.

\subsection{Equivalence of NGMI, ASI, and achievable FEC rate}
\label{sec:eq_metric}

Here we explain the equivalence of NGMI in \eqref{eq:NGMIps}, ASI in \eqref{eq:ASI}, and achievable FEC rate $R_{\text{fec}}^{*}$ in \eqref{eq:afr}. Our interests mainly lie in the equivalence under Gaussian channels with or without SNR mismatch between the true and auxiliary channels. 

\begin{theorem}
\label{theo:asi_mc}
Under an SNR-mismatched bitwise auxiliary channel, the ASI can be obtained with $s_{\text{o}}$ in \eqref{eq:Gauss_so}, $L_{\text{a}}$ in \eqref{eq:La}, and $p_{L_{\text{a}}}(l)$ in \eqref{eq:PLa} by Monte-Carlo integration as
\begin{IEEEeqnarray}{rCL}
	\label{eq:ASI_for_mc}
	\!\!\!\!\!\!\!\!\!\!\!\!\text{ASI} & = & 1 - \mathbb{E}_{L_{\text{a}}} \left[ \log_2 \left(1 + \exp \left( -\frac{s_{\text{o}}}{s}L_{\text{a}} \right) \right) \right]  \\
    	\!\!\!\!\!& = & 
    	\label{eq:asi_mc_2}
    	1 - \frac{1}{m} \sum_{i=1}^{m} \sum_{b\in \{0,1 \}} P_{B_i}(b) \cdot \nonumber \\
    	\!\!\!\!\! & &\!\!\!\!  \mathbb{E}_{\hat{L}_i^{\text{po}} \mid B_i = b} \left[ \log_2 \left(1+\exp \left( -(-1)^b \frac{s_{\text{o}}}{s} \hat{L}_i^{\text{po}}(\boldsymbol{Y}) \right) \right) \right] .
\end{IEEEeqnarray}
\end{theorem}

\begin{IEEEproof}
See Appendix A.
\end{IEEEproof}

\begin{theorem}
\label{theo:asi_eq_ngmi}
Under an SNR-mismatched bitwise auxiliary channel, 
\begin{IEEEeqnarray}{rCL}
	\label{eq:asi_gmis1}
	\text{ASI} = \text{NGMI} 
\end{IEEEeqnarray}
if and only if scaling parameters are $s = s_{\text{o}}$.
\end{theorem}

\begin{IEEEproof}
See Appendix B.
\end{IEEEproof}

\begin{theorem}
\label{theo:asi_afr}
Under an SNR-mismatched bitwise auxiliary channel,
\begin{IEEEeqnarray}{rCL}
	\label{eq:asi_afr}
	\text{ASI}  = \text{NGMI} = R_{\text{fec}}^{*} 
\end{IEEEeqnarray}
if and only if $s_{\rev{\text{d}}}=s_{\text{o}}/\rev{s}$.
\end{theorem}

\begin{IEEEproof}
Substituting $\mathbb{U}$ from \eqref{eq:unc_l} into \eqref{eq:afr} and setting $s_{\rev{\text{d}}} = s_\text{o}/s$ yields \eqref{eq:asi_mc_2}.
\end{IEEEproof}

In a deployable system, the L-values are quantized at the demapper, so the pmf of the quantized L-values is relevant to benchmark the performance. 
As explained in footnote~10 and Appendix~C, the ASI can be computed based on the quantized L-values just before an FEC decoder.
Before the quantization with a finite resolution ($n_{\text{L}}$), scaling of the L-values is important to determine the quantization step $\Delta l$. For example, a quantized L-value set  would be $\mathbb{L} = \{ \pm 0.5\Delta l , \pm 1.5\Delta l , \ldots , \pm l_{\text{max}} \}$, where $l_{\text{max}} = (n_{\text{L}}-1)\Delta l/2$. In a practical demapper, $L_i ({\boldsymbol{Y}})$ is scaled and rounded to the nearest element in the set $ \{ \pm 0.5, \pm 1.5 , \ldots , \pm l_{\text{max}} \}$. Here the scaling before the quantization in a deployable system detemines the quantization step $\Delta l$. Once L-values are quantized, the values represent just labels, so further scaling is not relevant unless the resolution is reduced.

Theorems~\ref{theo:asi_mc} to \ref{theo:asi_afr} are valid only when the L-values are not quantized, and Theorems~\ref{theo:asi_mc} and \ref{theo:asi_eq_ngmi} are valid only when the difference between the true and auxiliary channels are only SNR (e.g., both true and auxiliary channels are Gaussian channels with diffrent SNRs). Thus the rigorous equivalence will be lost in general fiber-optic channels with L-value quantization.
On the other hand, in research works for optical fiber communications under the discrete memoryless channel approximation with a reasonable quantization of L-values, the auxiliary channel can be very close to the true (approximated) channel. Thus we will see almost matched decoding and $s=s_{\text{o}}=s_{\rev{\text{d}}}=1$. As shown in Theorem~\ref{theo:asi_afr}, the NGMI, ASI, and $R_{\text{fec}}^{*}$ almost coincide.
As for the simulations in this work, we will derive matched \emph{a posteriori} L-values (so $s=s_{\text{o}}=s_{\rev{\text{d}}}=1$) in the following sections. This assumption is reasonable because in the following simulations we will compare the performance around the FEC limit, in which case we will see almost matched decoding also in deployable demapper assuming fixed $\widehat{\text{SNR}}$ (and thus $\text{NGMI} \approx \text{ASI} \approx R_{\text{fec}}^{*}$). In the rest of the paper we thus study performance under conditions where ASI, NGMI, and $R_{\text{fec}}^{*}$ are effectively equivalent, except where explicitly stated.

\section{Benchmarking in various bit mapping and modulation/shaping cases}
\label{sec:bl_mapg}
As explained in \cite[Remark~5]{bocherer_2019}, bit mapping will not change the performance metrics before the FEC decoding. On the other hand, it can influence \sout{to} the performance after practical FEC decoding.

While quaternary phase-shift keying (QPSK) has only a single bit per in-phase and quadrature component, high-order QAM has several. Therefore, an asymmetric L-value combination over bits for the codeword space, i.e., $[ L_{\text{a,1}}, L_{\text{a,2}}, \ldots , L_{\text{a},n} ]$ where $n$ denotes the FEC codeword length, becomes strongly dependent on the bit mapping for high-order QAM. The post-FEC decoding performance depends on the asymmetric L-value combination. The dependence is stronger in the case of a structured FEC code with a low code rate. 

The bit mapping over the codeword should be optimized for the used FEC, modulation, and shaping. In case of low-rate FEC for uniform signaling, the bit mapping affects the post-FEC BER even if the other conditions are the same \cite{hager_2014}, \cite[Fig.~5]{alvarado_ofc_2015}. Examples of good bit mappings for PAS were reported in \cite[Tab.~V]{bocherer_2015}.

\subsection{Considered bit mappings}
\label{sec:calc}
To estimate the information-theoretic quantities in Sec.~\ref{sec:theo}, 
Monte-Carlo integration is a useful way to approximate an integration by a finite sum. 
If $\boldsymbol{B}$ is represented by a sequence $\beta[\ell]$ and $\boldsymbol{L}$ by a sequence $\lambda[\ell]$ for $\ell=1,\ldots,m n_\text{s}$, where $n_\text{s}$ is the number of simulated symbols, then the symmetrized L-value sequence is calculated as $\lambda_\text{a}[\ell] = (-1)^{\beta[\ell]} \lambda[\ell]$ according to \eqref{eq:La}. Under a matched bitwise auxiliary channel (i.e., $s=s_{\text{o}}=s_{\rev{\text{d}}}=1$), the ASI \eqref{eq:ASI}, \eqref{eq:ASI_for_mc} can be computed with Monte-Carlo integration as
\begin{IEEEeqnarray}{rCL}
	\label{eq:mci_ASI}
	\text{ASI} \approx 1-\frac{1}{m n_{\text{s}}} \sum_{\ell=1}^{m n_{\text{s}}} \log_2 \left( 1 + \exp \left( - \lambda_{\text{a}}[\ell] \right) \right).
\end{IEEEeqnarray}
While a similar expression is found in \cite[Ths.~4.20,~4.21]{bicmbook}, \cite{alvarado_2015} for uniform signaling, it is valid for PAS, too, because the ASI works for both schemes as explained in Sec.~\ref{sec:rates}.\footnote{We believe that \eqref{eq:mci_ASI} is often used as NGMI in the optical fiber communication field because of its simplicity. Alternatively, one can compute the ASI by a discrete histogram of L-values fed into the FEC decoder. Details are provided in Appendix~C.}

Tab.~\ref{tab:sim_bit_mpg} shows the examined bit mappings. 
A bit mapping is denoted as $\mathbb{M}=[M_1,M_2,\ldots , M_n]$ for the bit indexes 1 to $n$ over the codeword, where each integer $1, \ldots , \bar{m}$ must occur $n/\bar{m}$ times in $\mathbb{M}$. Note that $\bar{m}=m/2$, i.e., the number of bits per PAM symbol for square QAM, and that $\bar{m}=m$ for cross QAM. To generate a square QAM symbol, two PAM symbols are combined. We consider the following three types of mappings.
\begin{itemize}
	\item \emph{Fixed structured bit mapping}: used for short-period structured bit mappings; $\mathbb{M}_{\text{FS1}}$ and $\mathbb{M}_{\text{FS2}}$ in Tab.~\ref{tab:sim_bit_mpg}. These bit mappings are practically implementable in today's DSP. 
	\item \emph{Random bit mapping}:
random permutation of bit position in a codeword by a random number, whose seed is changed at each codeword; $\mathbb{M}_{\text{R}}$ in Tab.~\ref{tab:sim_bit_mpg}. This is used only for benchmarking purposes.

	\item \emph{Fixed unstructured bit mapping}: random permutation of bit position in a codeword by a random number, whose seed is fixed; $\mathbb{M}_{\text{FU}}$ in Tab.~\ref{tab:sim_bit_mpg}. This is more practical than random bit mapping.
\end{itemize}
The first $k$ and the last $n-k$ components of $\mathbb{M}$ denote the bit tributaries of the information bits and parity bits, resp.
For square QAM, the bit tributaries of the corresponding one-dimensional symbol are shown, e.g., three bit tributaries for 64-QAM. Bit tributary 1 determines the sign of the one-dimensional symbol, and bit tributaries 2 and 3 determines the absolute amplitude of the one-dimensional symbol.
Usually we cannot control the distribution of parity bits, which are uniformly distributed. Thus not to change the pmf of PAS, we assign bit tributary 1 to the last $n/\bar{m}$ components in all considered bit mappings, where $n-k$ must be $\le n/\bar{m}$ for PAS.\footnote{Not all bits in bit tributary 1 are assigned to parity bits in the case of $n-k < n/\bar{m}$, but all bits in bit tributary 1 are placed in the last $n/\bar{m}$ components for simplicity.}

\begin{table}[t]
	\setlength{\unitlength}{.6mm} %
	\begin{center}
		\caption{Examined fixed structured ($\mathbb{M}_{\text{FS1}}$ and $\mathbb{M}_{\text{FS2}}$), random ($\mathbb{M}_{\text{R}}$), and fixed unstructured ($\mathbb{M}_{\text{FU}}$) bit mappings. 
				$\mathsf{R}(\mathbb{M},s_{\text{r}})$ denotes a random permutation of the vector $\mathbb{M}$ with the seed $s_{\text{r}}$.
				$\mathbb{M}^{\text{L}}$ denotes the first $n- n/\bar{m}$ components of $\mathbb{M}$.}
		\label{tab:sim_bit_mpg}
		\begin{tabular}{cc}
			\hline \hline
			Notation & Bit mapping \\ \hline
			$\mathbb{M}_{\text{FS1}}$ & $[\bar{m}, \ldots ,\bar{m}, \bar{m}-1, \ldots ,\bar{m}-1,\ldots , 1, \ldots , 1]$ \\
			$\mathbb{M}_{\text{FS2}}$ & $[\bar{m}, \bar{m}-1, \ldots , 2, \ldots , \bar{m}, \bar{m}-1, \ldots , 2, 1, \ldots , 1]$ \\ \hline
			$\mathbb{M}_{\text{R}}$ & $[\mathsf{R}(\mathbb{M}_{\text{FS1}}^{\text{L}}),\text{random}) , 1, \ldots , 1]$ \\ \hline
			$\mathbb{M}_{\text{FU}}$ & $[\mathsf{R}(\mathbb{M}_{\text{FS1}}^{\text{L}}),1), 1, \ldots , 1]$ \\
			\hline \hline
		\end{tabular}
	\end{center}
	\vspace{-0.4cm}
\end{table}

We employ the DVB-S2 low-density parity check (LDPC) code\cite{dvbs2} as an SD-FEC, whose FEC codeword length $n$ is 64800, and the number of decoding iterations was 50. The soft-demapping output is floating-point, and the decoding is done by floating-point operations with belief propagation. We assume the use of an outer HD-FEC to clean up the residual errors, having a code rate of 0.9922 and an input BER threshold of $5\times 10^{-5}$ \cite{millar_2015} for error-free operation.\footnote{The HD-FEC threshold is shown as the dashed line in the figures of post-FEC BER simulations.} In Fig.~\ref{fig:system}, the FEC encoder includes an outer systematic HD-FEC encoder, a time interleaver, and an SD-FEC encoder, with the reverse operations on the receiver side. Here we assume a sufficiently long temporal interleaving and no residual burst errors after the SD-FEC decoder (the same assumption is found in \cite[Sec.~II-B]{alvarado_2018_jlt}). If the interleaving length is insufficient, an error floor would result after HD-FEC decoding. The total FEC code rate is slightly smaller than that of the SD-FEC.

\subsection{Simulation results over the Gaussian channel under matched decoding}
\label{sec:sim_gc}

\begin{figure}[t]
	\begin{center}
		\setlength{\unitlength}{.6mm} %
		\scriptsize
		\includegraphics[scale=0.47]{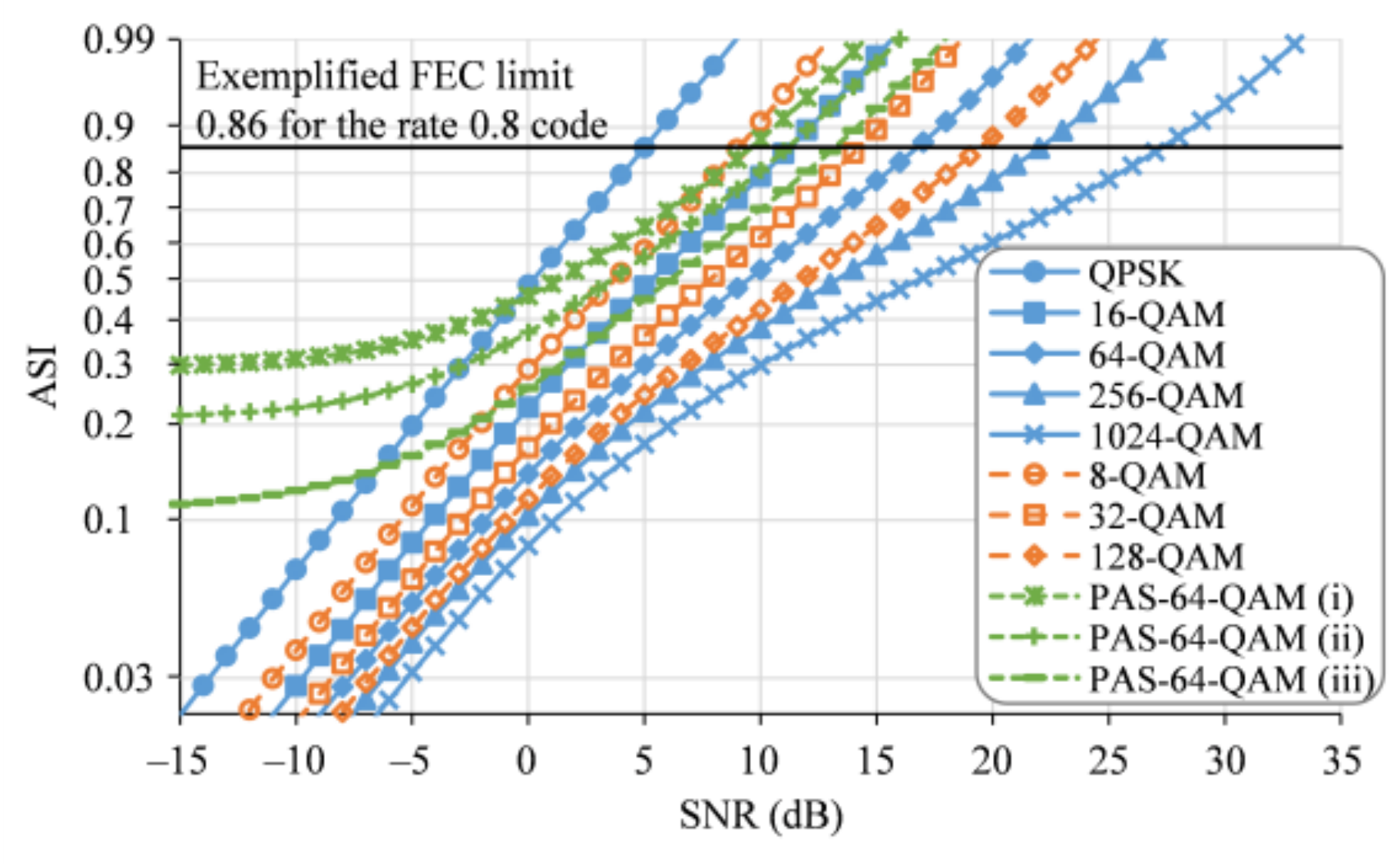}\\
		\vspace{-0.3cm}
		\caption{ASI as a function of SNR. The vertical axis is scaled according to $\log_{10} ( -\log_{10} (1-\text{ASI}))$ to make the QPSK-curve nearly linear. An alternative way to realize this linearization would be to plot $J^{-1}(\text{ASI})$, where $J(\cdot)$ denotes the J-function, used in \cite[Appendix]{tenbrink_2004}.}
		\label{fig:SNR_vs_metric}
	\end{center}
	\vspace{-0.3cm}
\end{figure}

Fig.~\ref{fig:SNR_vs_metric} shows the ASI from \eqref{eq:mci_ASI} for various modulation and shaping parameters as a function of SNR for the Gaussian channel under matched decoding, for which the ASI equals the NGMI as shown in Sec.~\ref{sec:eq_metric}.
Various modulation formats are examined and the parameters of PAS-64-QAM are shown in Tab.~\ref{tab:sim_8PAM}. The PAS scheme requires a code rate\footnote{Note that $m$ is defined as a two dimensional symbol in Sec.~\ref{sec:system}, so there are two sign bits.} $R_{\text{c}} \ge (m-2)/m$, so here the SD-FEC code rate $R_{\text{sdc}}$ is set to $5/6$ and bit mapping was set to $\mathbb{M}_{\text{FS1}}$.  
To generate PAS-64-QAM signals, constant composition distribution matching \cite{schulte_2016} was employed as the PS encoding with the PS codeword length of 1024 PAM symbols (512 QAM symbols). 
\begin{table}[t]
	\setlength{\unitlength}{.6mm} 
	\begin{center}
		\caption{One-dimensional pmf of PAS-64-QAM, and two-dimensional entropies and PS rate loss for simulations.}
		\label{tab:sim_8PAM}
		\begin{tabular}{ccccc}
			\hline \hline
			Parameters & Uniform & PAS (i) & PAS (ii) & PAS (iii) \\ \hline
			$P_{|X|}(1)$ & 0.250 & 0.698 & 0.611 & 0.494 \\
			$P_{|X|}(3)$ & 0.250 & 0.263 & 0.304 & 0.325 \\
			$P_{|X|}(5)$ & 0.250 & 0.037 & 0.075 & 0.141 \\
			$P_{|X|}(7)$ & 0.250 & 0.002 & 0.009 & 0.040 \\
			$\sum_{i=1}^{m}\mathbb{H}(B_i)$ & 6 & 4.238 & 4.754 & 5.356 \\
			$\mathbb{H}(\boldsymbol{B})$ & 6 & 4.124 & 4.604 & 5.226 \\
			$R_{\text{loss}}$ & 0 & 0.022 & 0.026 & 0.026 \\ \hline \hline
		\end{tabular}
	\end{center}
	\vspace{-0.4cm}
\end{table}
The performance metrics before FEC decoding does not depend on the bit mapping ($\mathbb{M}_{\text{FS1}}$, $\mathbb{M}_{\text{FS2}}$, $\mathbb{M}_{\text{R}}$, or $\mathbb{M}_{\text{FU}}$) in this simulation.
The required SNR at a given target ASI can be obtained from the plot, assuming no bit mapping dependence. For example, considering the nonideal performance of FEC, a suitable ASI is 0.86 for $R_{\text{c}}=0.8$ \cite[Tab.~II]{koike_jlt_2017} (see also footnote~\ref{ft:codegap}).
The ASI for PAS shows a floor and does not approach zero at SNR:s $<-15$ dB because of the nonuniformity. Note that no PAS-64-QAM system is available for reliable communications in this low-SNR regime, because the FEC code rate cannot be equal to the ASI or lower without violating the constraint $R_\text{c} \ge (m-2)/m$.

\begin{figure}[t]
	\begin{center}
		\setlength{\unitlength}{.6mm} %
		\scriptsize
		\includegraphics[scale=0.47]{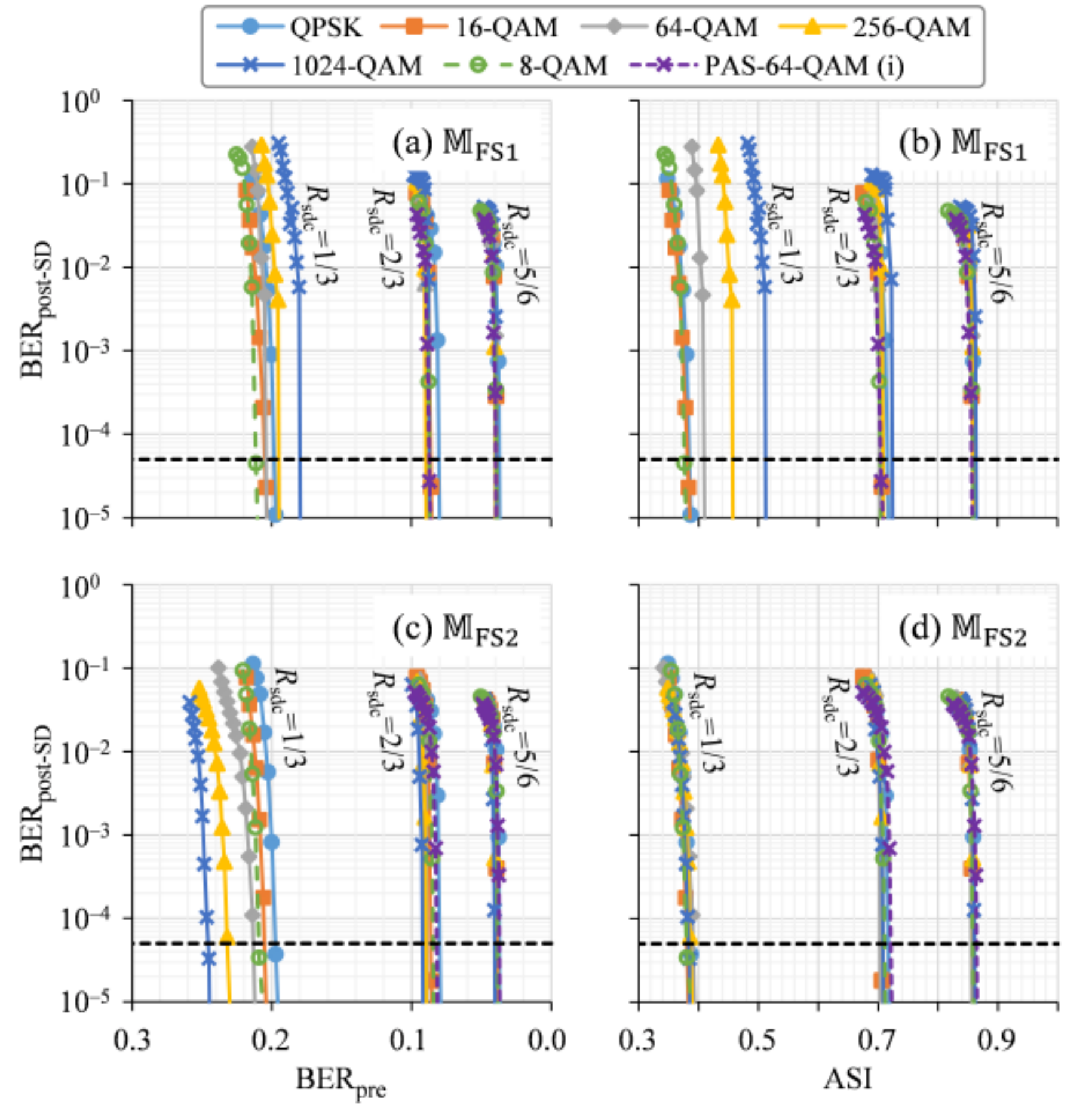}\\
		\vspace{-0.3cm}
		\caption{Post-SD-FEC BER with fixed structured bit mappings. (a) and (b) are with $\mathbb{M}_{\text{FS1}}$, (c) and (d) are with $\mathbb{M}_{\text{FS2}}$. The dependence on pre-FEC BER is shown in (a) and (c), and on ASI in (b) and (d).}
		\label{fig:BERpost_FS}
	\end{center}
	\vspace{-0.3cm}
\end{figure}

\begin{figure}[t]
	\begin{center}
		\setlength{\unitlength}{.6mm} %
		\scriptsize
		\includegraphics[scale=0.47]{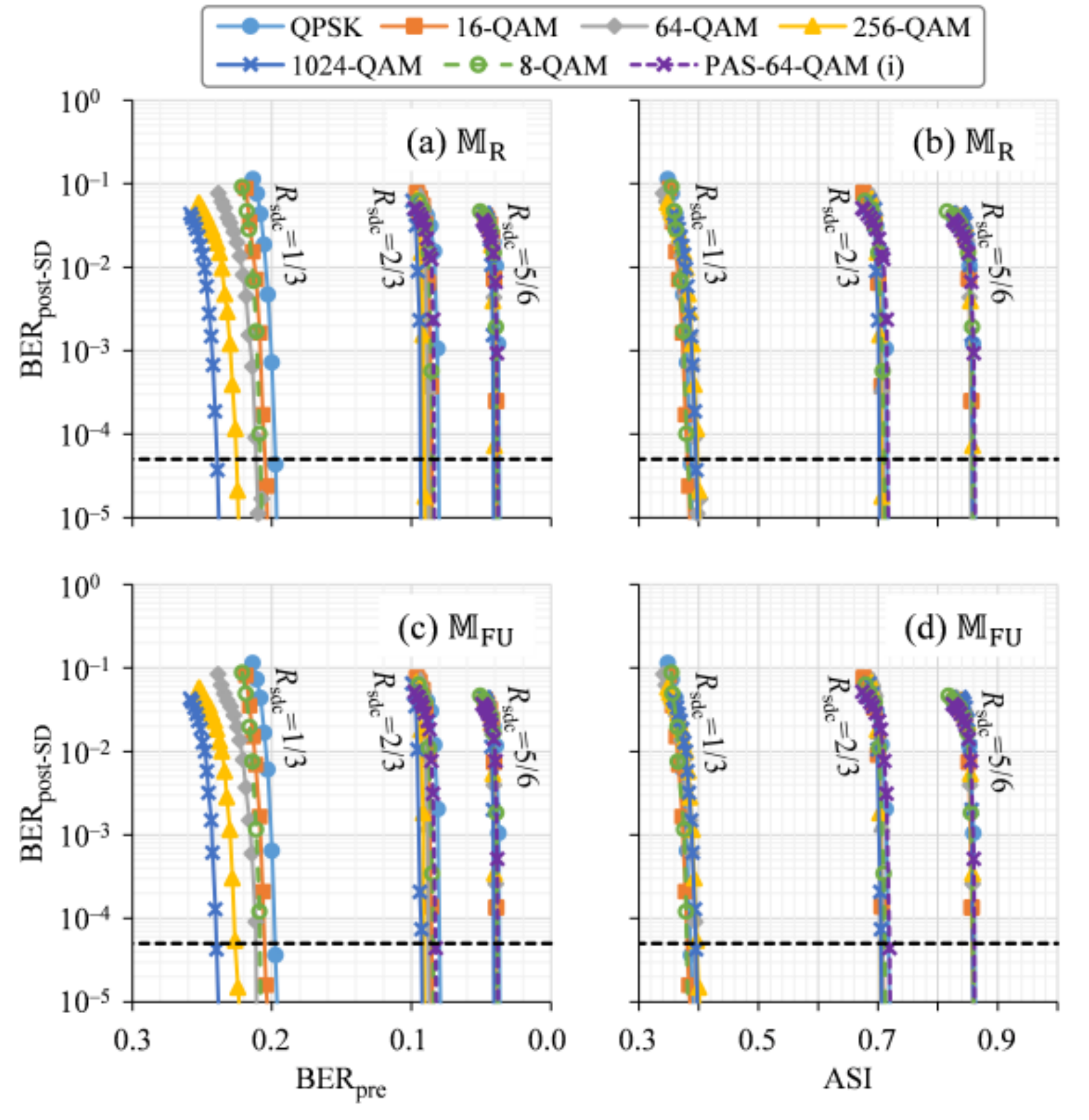}\\
		\vspace{-0.3cm}
		\caption{Post-SD-FEC BER with random ($\mathbb{M}_{\text{R}}$) or fixed unstructured ($\mathbb{M}_{\text{FU}}$) bit mappings. (a) and (b) are with $\mathbb{M}_{\text{R}}$, (c) and (d) are with $\mathbb{M}_{\text{FU}}$. The dependence on pre-FEC BER is shown in (a) and (c), and on ASI in (b) and (d).}
		\label{fig:BERpost_R}
	\end{center}
	\vspace{-0.3cm}
\end{figure}

Figs.~\ref{fig:BERpost_FS} and \ref{fig:BERpost_R} summarize the post-SD-FEC BER $\text{BER}_{\text{post-SD}}$, i.e., the BER after SD-FEC decoding, as a function of the pre-FEC BER defined in \eqref{eq:preBER}, and ASI).
The modulation formats are Gray-coded QPSK, 16-QAM, 64-QAM, 256-QAM, and 1024-QAM, star-8-QAM labeled as in Tab.~\ref{tab:sym_mapg_8qam}, and PAS-64-QAM (i) in Tab.~\ref{tab:sim_8PAM}. The SD-FEC code rates $R_{\text{sdc}}$ are $1/3$, $2/3$, and $5/6$ 
uniform QAM and $2/3$ and $5/6$ for the PAS. 
\begin{table}[t]

	\caption{Mapping of bits $\boldsymbol{B}=B_1 B_2 B_3$ to a star-8-QAM symbol $\boldsymbol{X}$.}
	\label{tab:sym_mapg_8qam}
	\vspace{-0.4cm}
	\begin{center}
	\begin{tabular}{|cc|cc|cc|cc|}
		\hline
		$\boldsymbol{B}$ & $\boldsymbol{X}$ & $\boldsymbol{B}$ & $\boldsymbol{X}$ \\
		\hline\hline
		000 & $[ 1+\sqrt{3}, 1+\sqrt{3}]$ & 110 & $[-1-\sqrt{3}, -1-\sqrt{3}]$ \\
		001 & $[            0,            2]$ & 111 & $[            0,            -2]$ \\
		011 & $[-1-\sqrt{3}, 1+\sqrt{3}]$ & 101 & $[ 1+\sqrt{3}, -1-\sqrt{3}]$ \\
		010 & $[           -2,            0]$ & 100 & $[            2,            0]$ \\
		\hline
	\end{tabular}
	\end{center}
\end{table}

In the case of fixed structured bit mappings in Fig.~\ref{fig:BERpost_FS}, the post-SD-FEC BER of the signal with high-order modulation and low code rates strongly depends on the bit mapping. This dependence comes from the structure of the code, so this is a limitation of using code-independent performance metrics.
The bit mapping $\mathbb{M}_{\text{FS2}}$ shows better post-FEC BER benchmark performance than $\mathbb{M}_{\text{FS1}}$ based on the ASI, while benchmarks based on the pre-FEC BER are inaccurate at the FEC code rate $R_{\text{sdc}}=1/3$ for both mappings. 
The BER benchmarking can fail if the fixed structured bit mapping is not tailored for the code, so the post-FEC BER does indeed depend on the combination of the FEC code structure and bit mapping.

As for the post-SD-FEC BER curves with a random bit mapping $\mathbb{M}_{\text{R}}$, shown in the upper side of Fig.~\ref{fig:BERpost_R}, ASI is clearly a better metric than pre-FEC BER because the convergence of the curves is significantly better. 
This random bit mapping breaks the memory in the practical codes\footnote{The memory break for PAS is limited because the last part of the $\mathbb{M}_{\text{R}}$ is $[1,\ldots,1]$.} 
However, random bit mapping is too complex to be practical with today's DSP and is therefore only used for benchmarking. 
Instead the \emph{fixed unstructured bit mapping} can be an interesting practical choice. The curves in the lower part of Fig.~\ref{fig:BERpost_R} show almost the same behavior as for random bit mapping. These results were partly expected, but was not fully predictive in a practical FEC decoding before this examination.
The curves of the post-SD-FEC BER vs. ASI for uniform signaling with $\mathbb{M}_{\text{FU}}$ are converged. We examined nine other random seeds for the fixed unstructured bit mapping, and we found no significant difference in the results.
While $\mathbb{M}_{\text{FS1}}$ fails the benchmark at $R_{\text{sdc}}=1/3$, it shows smaller required SNR than the others for PAS-64-QAM by 0.1--0.2 dB at $R_{\text{sdc}}=5/6$ or 0.3 dB at $R_{\text{sdc}}=2/3$ though the small differences are not clearly visible in Figs.~\ref{fig:BERpost_FS} and \ref{fig:BERpost_R}. 
For the all considered bit mappings, ASI is a suitable benchmark of the post-FEC BER, especially for PAS.

\section{Benchmark for PAS-QAM over fiber-optic channel}
\label{sec:fo_ch}
In this section, we discuss the benchmark accuracy of the pre-FEC BER in \eqref{eq:preBER}, normalized AIR in \eqref{eq:RBMD_HB}, and ASI in \eqref{eq:mci_ASI}.
The simulation parameters are listed in Tab.~\ref{tab:sim_NL}. The split-step Fourier method with the Manakov equations was used for simulating the signal propagation over fibers. Lumped optical amplification is used to compensate for the loss of the fiber, and the amplified spontaneous emission noise was loaded per span. 
In the receiver, the residual chromatic dispersion was compensated in the frequency domain. Adaptive equalization and carrier recovery were performed by fully pilot-aided signal processing\cite{mikael_2019_oe}. The number of taps in the adaptive equalizer was 21 at 2 samples per symbol, and the adaptation was done by the constant modulus algorithm based on QPSK pilots. 
The carrier phase at a data symbol was computed by a linear interpolation of the estimated phases from the neighboring two pilot symbols, which were recovered by moving average over the previous and the next pilot symbols. 
In the soft demapping, the SNR for the auxiliary channel (approximated by the discrete memoryless Gaussian channel) was estimated from the pilot signals. The used FEC was described in Sec.~\ref{sec:bl_mapg}. 
The resolution and effective number of bits in digital-to-analog and analog-to-digital conversions were 8 and 6 bits, resp.\footnote{The minor differences in the simulation parameters compared to our previous work \cite{yoshida_ecoc_2017} are the target pmf (entropies of a wider range are studied here), the quantization of L-values ($2^5\to2^{11}$), the decoding iterations ($20\to50$), symbol length ($\sim2^{15}\to{}\sim2^{16}$), number of spans ($5\to1\sim50$), and codewords per waveform ($5\to11$). }

\begin{table}[t]
	\setlength{\unitlength}{.6mm} %
	\begin{center}
		\caption{Simulation parameters.}
		\label{tab:sim_NL}
		\begin{tabular}{ll}
			\hline \hline
			Parameter & Value \\ \hline
			Symbol rate & $32$ Gsymbol/s \\
			Spectral shaping & root-raised cosine, $1\%$ roll-off \\
			Channel spacing & $32 \times 1.01$ GHz \\
			Number of channels & $7$ \\
			Launch power & $[-7.5, +4.5]$ dBm/channel \\
			Fiber attenuation & $0.2$ dB/km \\
			Chromatic dispersion & $17$ ps/nm/km \\
			Nonlinear coefficient & $1.2$ $\text{W}^{-1}\text{km}^{-1}$ \\
			Span length & $100$ km \\
			Number of spans & $[1, 50]$ (PAS-64-QAMs) \\
			& $[1, 30]$ Uniform 64-QAM) \\
			Noise figure of amplifier & $5$ dB \\
			Laser linewidth & $100$ kHz \\
			Pilot insertion ratio & $4\%$ \\
			\hline \hline
		\end{tabular}
	\end{center}
	\vspace{-0.4cm}
\end{table}

To simulate lower post-FEC BER from nonlinearly propagated signals with limited simulation resources, we employed the virtual interleave and scramble technique presented in \cite[Sec.~VII-F]{bocherer_2019}, \cite{schmalen_2012,stojanovic_2013,buchali_2016,yoshida_ofc18_tool}.  
Then, totally more than 500 codewords were simulated to obtain each point. The bit mappings $\mathbb{M}_{\text{FS1}}$ and $\mathbb{M}_{\text{FU}}$ were examined. Each codeword was further interleaved over time, both polarizations, and both quadratures.

\begin{figure}[t]
	\begin{center}
		\setlength{\unitlength}{.6mm} %
		\scriptsize
		\includegraphics[scale=0.47]{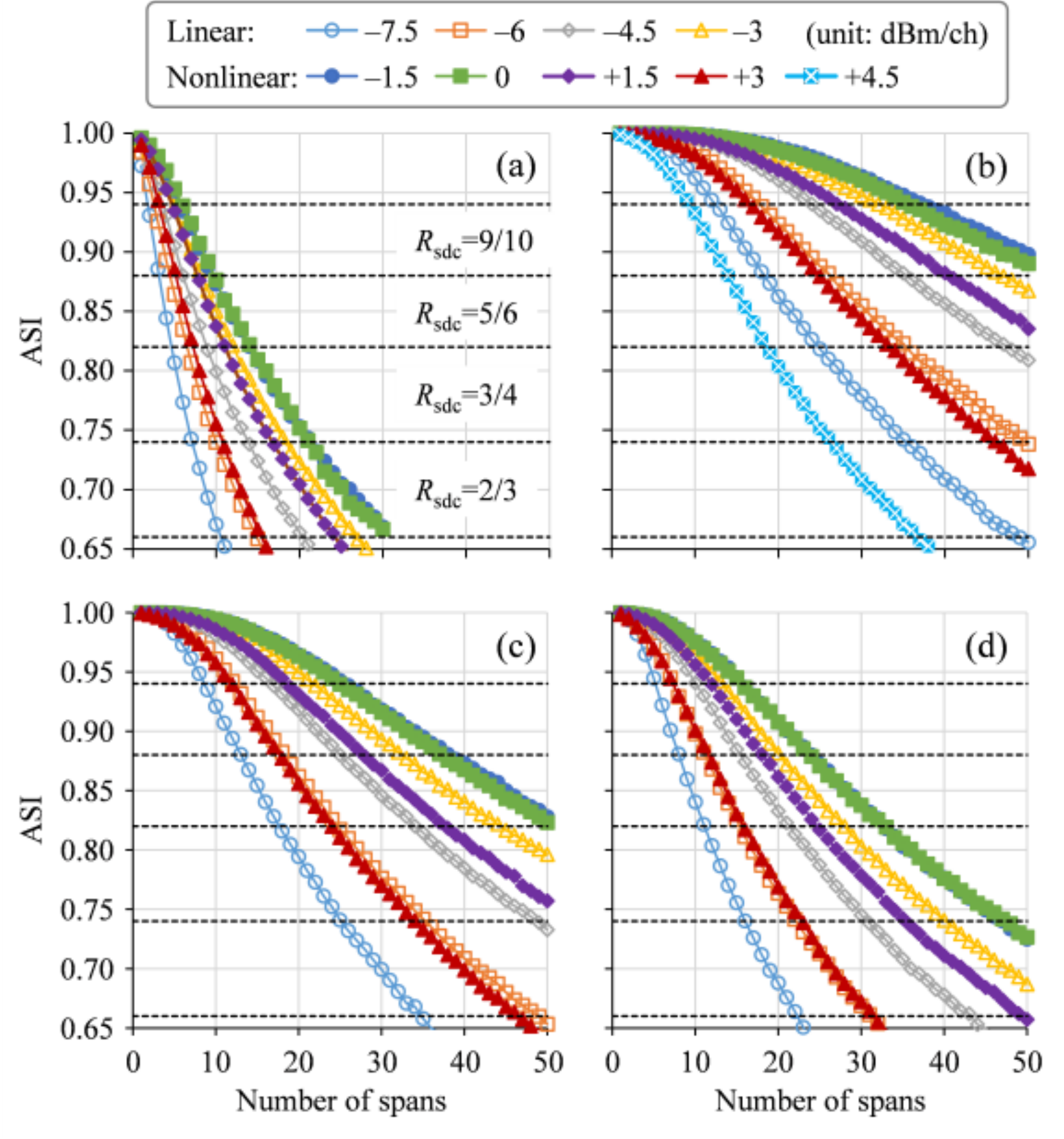}\\
		\vspace{-0.3cm}
		\caption{ASI vs. number of spans for (a) uniform 64-QAM, (b) PAS-64-QAM (i), (c) PAS-64-QAM (ii), and (d) PAS-64-QAM (iii).}
		\label{fig:NLsim}
	\end{center}
	\vspace{-0.3cm}
\end{figure}

Fig.~\ref{fig:NLsim} shows ASI vs. number of spans for (a) uniform 64-QAM, (b) PAS-64-QAM (i), (c) PAS-64-QAM (ii), and (d) PAS-64-QAM (iii). FEC decoding with an SD-FEC code rate $R_{\text{sdc}}$ 2/3, 3/4, 5/6, or 9/10 was performed for ASI = $[0.66, 0.74]$, $[0.74, 0.82]$, $[0.82, 0.88]$, or $[0.88, 0.94]$. To include both linear and nonlinear cases for each shaping parameter and each code rate, we studied 4.5 dBm/channel for PAS-64-QAM (i). Except for that, the maximum launch power was 3 dBm/channel.

For a given combination of number of spans and launch power, we can choose suitable shaping and FEC parameters from Fig.~\ref{fig:NLsim}. For example, at 25 spans and 0 dBm/channel, PAS-64-QAM (ii) with $R_{\text{sdc}}=9/10$ or PAS-64-QAM (iii) with $R_{\text{sdc}}=5/6$ would be good candidates for reliable and spectrally efficient communication.

\begin{figure}[t]
	\begin{center}
		\setlength{\unitlength}{.6mm} %
		\scriptsize
		\includegraphics[scale=0.5]{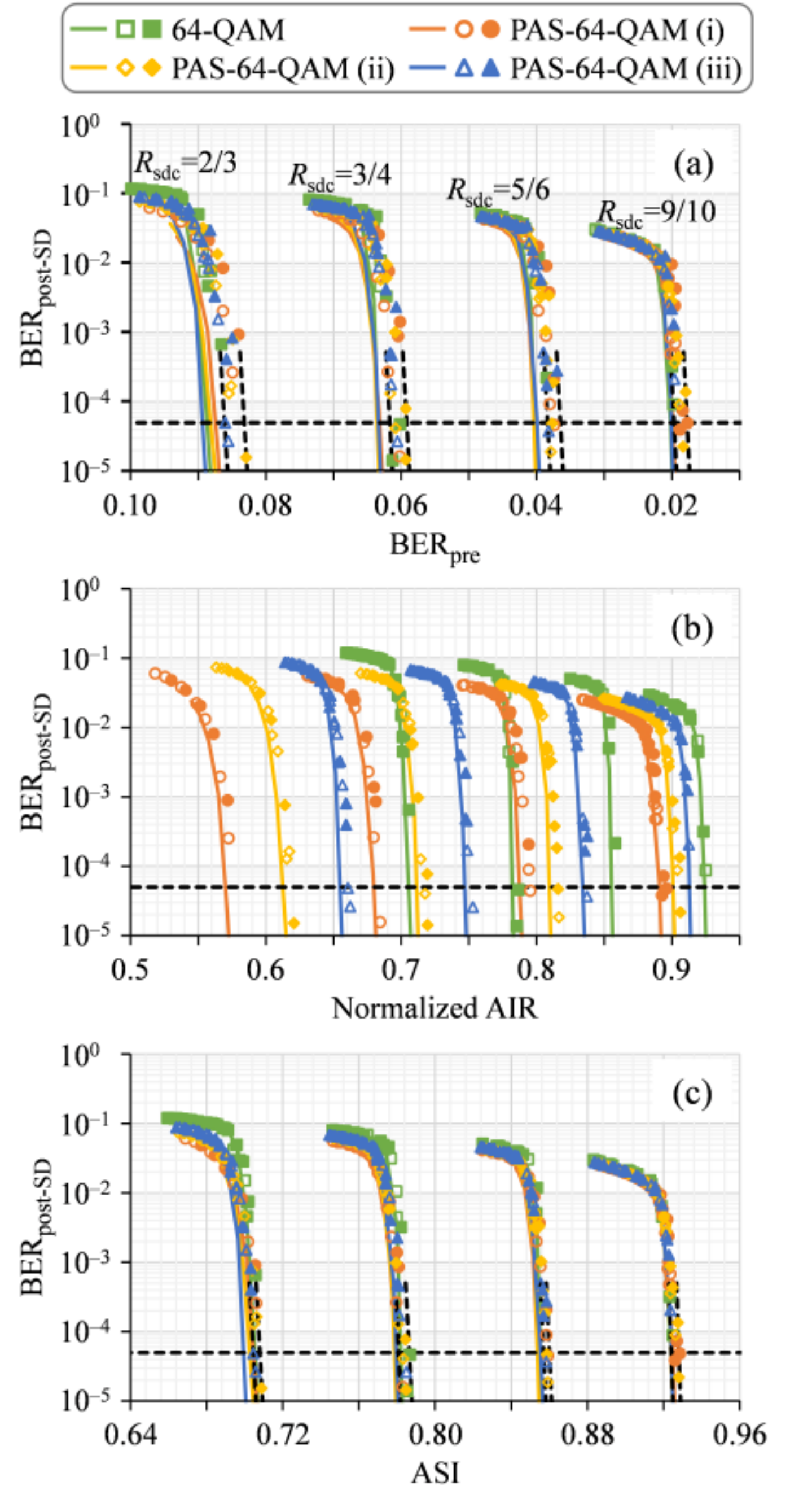}\\
		\vspace{-0.3cm}
		\caption{Post-SD-FEC BER as a function of (a) pre-FEC BER, (b) normalized AIR, or (c) ASI on fiber-optic channel with bit mapping $\mathbb{M}_{\text{FS1}}$.}
		\label{fig:BERpostNL_Pi0}
	\end{center}
	\vspace{-0.3cm}
\end{figure}

\begin{figure}[t]
	\begin{center}
		\setlength{\unitlength}{.6mm} %
		\scriptsize
		\includegraphics[scale=0.5]{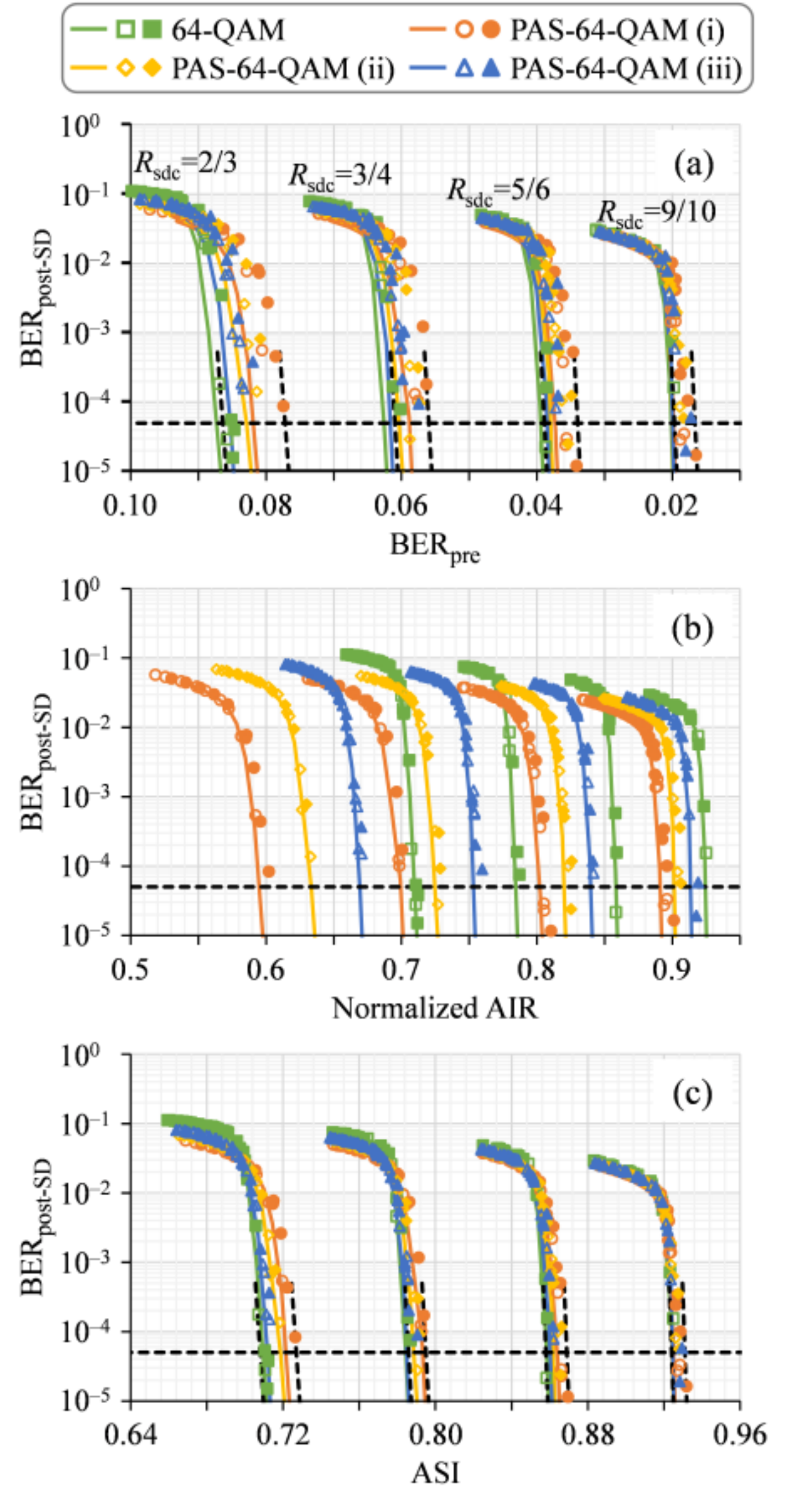}\\
		\vspace{-0.3cm}
		\caption{Post-SD-FEC BER as a function of (a) pre-FEC BER, (b) normalized AIR, or (c) ASI on fiber-optic channel with bit mapping $\mathbb{M}_{\text{FU}}$.}
		\label{fig:BERpostNL_Pi3}
	\end{center}
	\vspace{-0.3cm}
\end{figure}

Figs.~\ref{fig:BERpostNL_Pi0} and \ref{fig:BERpostNL_Pi3} show post-SD-FEC BER as a function of the various metrics (pre-FEC BER, normalized AIR, ASI). Open markers show the low nonlinearity cases with launch powers of $[-7.5, -3]$ dBm/channel, and filled markers those with high nonlinearity, $[-1.5, +4.5]$ dBm/channel. For reference, simulation results over the Gaussian channel are shown by solid lines. The vertical dashed lines in Figs.~\ref{fig:BERpostNL_Pi0}(a), \ref{fig:BERpostNL_Pi0}(c), \ref{fig:BERpostNL_Pi3}(a), and \ref{fig:BERpostNL_Pi3}(c) show the peak-to-peak variation of the curves in the nonlinear transmission cases.
At the same FEC code rate, shaping parameter, and a given post-FEC BER, relatively larger pre-FEC BER or almost the same ASI is required in the fiber-optic channel compared with the Gaussian channel. Normalized AIR does not work, and in particluar low-entropy cases (stronger shaping) are problematic. The convergence of the curves with $\mathbb{M}_{\text{FS1}}$ is worse than that with $\mathbb{M}_{\text{FU}}$. The benchmark errors are summarized in Tab.~\ref{tab:predict_sum}. The errors at a given $R_{\text{sdc}}$ are quantified both horizontally ($\Delta \text{Metric}$) and vertically ($\Delta \text{BER}_{\text{post-SD}}$) around the post-SD-FEC BER of $5\times 10^{-5}$, which corresponds to a certain HD-FEC limit \cite{millar_2015}. The benchmark error $\Delta \text{BER}_{\text{pre}}$ appears to be smaller than $\Delta \text{ASI}$, but since its variation is around a much smaller metric value, the relative error is larger. This observation is consistent with $\Delta \text{BER}_{\text{post-SD}}$, cf.~\cite{yoshida_ptl_2017,yoshida_ecoc_2017}, whose variation with ASI is between 4 and 790 times less than the corresponding variation with pre-FEC BER for $\mathbb{M}_{\text{FS1}}$, and between 16 and 57,000 times less for $\mathbb{M}_{\text{FU}}$.

\begin{table}[t]
	\setlength{\unitlength}{.6mm} %
	\begin{center}
		\caption{Simulated benchmark errors of post-SD-FEC BER over the fiber-optic channels.}
		\label{tab:predict_sum}
		\begin{tabular}{|c|cc|cc|}
			\hline
			\multicolumn{5}{|c|}{Bit mapping: $\mathbb{M}_{\text{FS1}}$} \\
			\hline
			\multicolumn{1}{|c|}{$R_{\text{sdc}}$} & \multicolumn{2}{c}{$\Delta \text{Metric}$} & \multicolumn{2}{|c|}{$\Delta \text{BER}_{\text{post-SD}}$} \\
			 & $\text{BER}_\text{pre}$ & ASI & $\text{BER}_\text{pre}$ & ASI \\
			\hline
			2/3 & $2.9 \times 10^{-3}$ & $3.5 \times 10^{-3}$ & $4.4 \times 10^{4}$ & $5.6 \times 10^{1}$ \\
			3/4 & $2.5 \times 10^{-3}$ & $5.0 \times 10^{-3}$ & $3.2 \times 10^{4}$ & $3.2 \times 10^{2}$ \\
			5/6 & $1.9 \times 10^{-3}$ & $3.0 \times 10^{-3}$ & $2.6 \times 10^{3}$ & $6.3 \times 10^{1}$ \\
			9/10 & $1.9 \times 10^{-3}$ & $4.0 \times 10^{-3}$ & $6.3 \times 10^{3}$ & $1.6 \times 10^{3}$ \\
			\hline\hline
			\multicolumn{5}{|c|}{Bit mapping: $\mathbb{M}_{\text{FU}}$} \\
			\hline
			\multicolumn{1}{|c|}{$R_{\text{sdc}}$} & \multicolumn{2}{c}{$\Delta \text{Metric}$} & \multicolumn{2}{|c|}{$\Delta \text{BER}_{\text{post-SD}}$} \\
			 & $\text{BER}_\text{pre}$ & ASI & $\text{BER}_\text{pre}$ & ASI \\
			\hline
			2/3 & $9.2 \times 10^{-3}$ & $1.8 \times 10^{-2}$ & $9.1 \times 10^{11}$ & $1.6 \times 10^{7}$ \\
			3/4 & $5.0 \times 10^{-3}$ & $8.0 \times 10^{-3}$ & $1.0 \times 10^{8}$ & $1.0 \times 10^{4}$ \\
			5/6 & $4.9 \times 10^{-3}$ & $1.1 \times 10^{-2}$ & $2.0 \times 10^{9}$ & $1.4 \times 10^{7}$ \\
			9/10 & $3.0 \times 10^{-3}$ & $6.0 \times 10^{-3}$ & $1.0 \times 10^{6}$ & $6.3 \times 10^{4}$ \\
			\hline
		\end{tabular}
	\end{center}
	\vspace{-0.4cm}
\end{table}

From the Gaussian channel simulations over a wide range of spectral efficiencies in Figs.~\ref{fig:BERpost_FS} and \ref{fig:BERpost_R}, the ASI benchmarks post-FEC BER better than pre-FEC BER in the case of \emph{fixed unstructured bit mapping} $\mathbb{M}_{\text{FU}}$.  
On the other hand, as explained in Sec.~\ref{sec:sim_gc}, last paragraph, $\mathbb{M}_{\text{FS1}}$ gives a slightly smaller ASI (smaller SNR) at an FEC threshold than $\mathbb{M}_\text{FU}$ for PAS-64-QAMs (and also for uniform 64-QAM). 
In the transmission simulation in Figs.~\ref{fig:BERpostNL_Pi0} and \ref{fig:BERpostNL_Pi3} with a limited range of spectral efficiencies,  
$\mathbb{M}_{\text{FS1}}$ shows not only better performance at an FEC threshold but also better accuracy of post-FEC BER benchmarking than $\mathbb{M}_\text{FU}$ under the (admittedly limited) examined modulation and FEC cases. 

Here we can suggest an open problem for possible future research.
We start with finding a good (nearly optimum) bit mapping $\mathbb{M}_{\text{o}}$, which can be different for each combination of modulation, shaping, and FEC code rate. Then, we quantify numerically the relation between the ASI and post-FEC BER. At least the required SNR performance will be nearly optimum for each signal. Here it would be very interesting to study how the benchmarking accuracy of also the post-FEC BER becomes. Judging from the results in this section, the post-FEC benchmarking accuracy might be good as well. A good bit mapping slightly improves the nonideal performance of a practical FEC decoder, which cannot be taken into account in performance metrics that usually benchmark a post-FEC BER with an ideal FEC.

\section{Conclusions}
\label{sec:cncl}
For the first time, we defined the generalized L-values under mismatched decoding and L-value scaling, and summarized relevant performance metrics to benchmark the post-FEC BER. We then showed theoretically the equivalence of NGMI, ASI, and achievable FEC rate $R_{\text{fec}}^{*}$ for matched decoding in a bitwise receiver without quantization and scaling of L-values. 
We then examined, for approximately matched decoding, where $\text{ASI} \approx \text{NGMI} \approx R_{\text{fec}}^{*}$, the metrics to benchmark the post-SD-FEC BER under various combinations of modulation, shaping, and code rates. The limitation of the metric is the post-SD-FEC BER dependence on the code structure in a practical non-ideal FEC. 
In the case of high code rates, e.g., $\ge 2/3$, there are less issues, and even the pre-FEC BER would work, e.g., for in-service performance monitoring, if the FEC limit difference is taken into account.
The post-SD-FEC BER at lower code rates depends on the bit mapping more significantly. Thus, the ASI,  NGMI, and $R_{\text{fec}}^{*}$, which cannot take the non-ideal FEC performance into account, do not always benchmark the post-SD-FEC BER accurately. However, when we break the bit mapping dependence by a random bit mapping, or even fixed unstructured bit mapping, the benchmarking accuracy improves significantly.  
The benchmarking by the pre-FEC BER is significantly worse at low code rates even if we \rev{employ} a random or fixed unstructured bit mapping. 
The observation for the Gaussian channel is useful for the fiber-optic channel with nonlinearity as well, though the benchmarking errors become larger for limited data sizes or non-Gaussian-channel signal degradations. 

\section*{Acknowledgments}
We thank Koji Igarashi of Osaka University for fruitful discussions about performance metrics.

\section*{Appendix A \\ Proof of Theorem 1}

A reformulated description of ASI in \eqref{eq:ASI} is
\begin{IEEEeqnarray}{rCL}
	\label{eq:ASI_MI}
	\text{ASI} &=& \int_{-\infty}^{\infty} p_{L_{\text{a}}}(l) \log_2 \frac{2 p_{L_{\text{a}}}(l)}{q_{| L_{\text{a}} |}(| l |)} \text{d} l , \\
	\label{eq:ASI_for_mc0}
	 &=& 1 - \int_{-\infty}^{\infty} p_{L_{\text{a}}}(l) \log_2 \left(1 + \frac{p_{L_{\text{a}}}(-l)}{p_{L_{\text{a}}}(l)}\right) \text{d} l . 
\end{IEEEeqnarray}
An extrinsic L-value has an inherent property of \emph{consistency} \cite[Th.~3.10]{bicmbook}
\begin{IEEEeqnarray}{rCL}
	\label{eq:Lex_con}
	\frac{p_{L_i^{\text{ex}} \mid B_i}(l \! \mid \! 0)}{p_{L_i^{\text{ex}} \mid B_i}(l \! \mid \! 1)} & = & \text{e}^l 
\end{IEEEeqnarray}
if the extrinsic L-values are computed from the true biwise channel $p_{\boldsymbol{Y} | B_i}(\boldsymbol{y} | b_i)$, i.e., 
\begin{IEEEeqnarray}{rCL}
	\label{eq:Lex_true}
	L_{i}^{\text{ex}}(\boldsymbol{y}) & = & \frac{p_{\boldsymbol{Y} \mid B_i}(\boldsymbol{y} \! \mid \! 0)}{p_{\boldsymbol{Y} \mid B_i}(\boldsymbol{y} \! \mid \! 1) } .
\end{IEEEeqnarray}
From \eqref{eq:Lex_con}, a fraction of pdfs of true \emph{a posteriori} L-values is given by
\begin{IEEEeqnarray}{C}
	\label{eq:Lpo_con}
	\frac{p_{L_i^{\text{po}} \mid B_i}(l \! \mid \! 0)}{p_{L_i^{\text{po}} \mid B_i}(l \! \mid \! 1)}  = \text{e}^{l-L_i^{\text{pr}}} .
\end{IEEEeqnarray}
Then, a fraction of pdfs of L-values at the soft-demapping output is given by
\begin{IEEEeqnarray}{C}
	\label{eq:qLpo_con}
	\!\!\!\!\!\!\!\!\!\!  \frac{p_{L_i \mid B_i}(l \! \mid \! 0)}{p_{L_i \mid B_i}(l \! \mid \! 1)}  =
	\frac{p_{\hat{L}_i^{\text{po}} \mid B_i}(l \! \mid \! 0)}{p_{\hat{L}_i^{\text{po}} \mid B_i}(l \! \mid \! 1)}  = \frac{p_{L_i^{\text{po}} \mid B_i}( \frac{s_{\text{o}}}{s} l \! \mid \! 0)}{p_{L_i^{\text{po}} \mid B_i}( \frac{s_{\text{o}}}{s} l \! \mid \! 1)} = \text{e}^{ \frac{s_{\text{o}}}{s} l - L_i^{\text{pr}}} .
\end{IEEEeqnarray}
The expression in \eqref{eq:qLpo_con} is then reformulated to
\begin{IEEEeqnarray}{rCL}
	\label{eq:sLapo_con}
	p_{L_i \mid B_i}(l \! \mid \! 0) & = & \text{e}^{\frac{s_{\text{o}}}{s}l} \frac{P_{B_i}(1)}{P_{B_i}(0)} p_{L_i \mid B_i}(l \! \mid \! 1).
\end{IEEEeqnarray}
The pdf of asymmetric L-value per bit tributary $L_{\text{a},i}=(-1)^{B_i} L_i(\boldsymbol{Y})$ is
\begin{IEEEeqnarray}{rCL}
	\label{eq:Lai_con}
	p_{L_{\text{a},i}}(l) & = & \sum_{b\in\mathbb{B}} P_{B_{i}}(b) p_{L_i \mid B_i}( (-1)^{b} l  \! \mid \! b) .
\end{IEEEeqnarray}
Using \eqref{eq:sLapo_con} in \eqref{eq:Lai_con} gives
\begin{IEEEeqnarray}{rCL}
	\label{eq:Lai_con2}
	\!\!\!\!\!\!\!\! p_{L_{\text{a},i}}(l)  & = & P_{B_{i}}(0) \text{e}^{\frac{s_{\text{o}}}{s}l} \frac{P_{B_i}(1)}{P_{B_i}(0)} p_{L_i \mid B_i}(l \! \mid \! 1)  \nonumber \\
	\!\!\!\!\!\!\!\! &&  + P_{B_i}(1) p_{L_i \mid B_i}(-l \! \mid \! 1) \\
	\label{eq:Lai_pdf}
	\!\!\!\!\!\!\!\! & = & P_{B_i}(1) \! \left( \text{e}^{\frac{s_{\text{o}}}{s}l} p_{L_i \mid B_i}(l \! \mid \! 1) + p_{L_i \mid B_i}(-l \! \mid \! 1)\right) .
\end{IEEEeqnarray}
The pdf of $L_{\text{a}}$ is
\begin{IEEEeqnarray}{rCL}
	\!\!\!\!p_{L_{\text{a}}}(l) & = & \frac{1}{m} \sum_{i=1}^m p_{L_{\text{a},i}}(l)  \\
	\label{eq:La_pdf}
	& = & \frac{1}{m} \sum_{i=1}^m  \left( \text{e}^{\frac{s_{\text{o}}}{s}l} p_{B_i, L_i}(1, l) + p_{B_I, L_i}(1, -l)\right) . 
\end{IEEEeqnarray}
From \eqref{eq:La_pdf} we obtain
\begin{IEEEeqnarray}{rCL}
	\label{eq:La_ratio}
	\frac{p_{L_{\text{a}}}(-l)}{p_{L_{\text{a}}}(l)} & = & \text{e}^{-\frac{s_{\text{o}}}{s}l} .
\end{IEEEeqnarray}
Applying \eqref{eq:La_ratio} to \eqref{eq:ASI_for_mc0}, \eqref{eq:ASI_for_mc} is derived, which completes the proof.

\section*{Appendix B \\ Proof of Theorem 2}
The $I_{q,s}^{\text{gmi}} (\boldsymbol{B}; \boldsymbol{Y})$ is reformulated from \eqref{eq:GMIs} and \eqref{eq:aux_chs}, as well as the reformulation in \cite[Th.~4.11]{bicmbook},
\begin{IEEEeqnarray}{l}
	\label{eq:proof2}
	I_{q,s}^{\text{gmi}} (\boldsymbol{B}; \boldsymbol{Y}) \nonumber \\
	\,\,\,\, =  \mathbb{E}_{\boldsymbol{B},\boldsymbol{Y}} \!\! \left[  \log_2 \frac{ \frac{\Pi_{i=1}^m P_{B_i}(B_{i})}{P_{\boldsymbol{B}}(\boldsymbol{B})} \Pi_{i=1}^m q_{\boldsymbol{Y} \mid B_i} (\boldsymbol{Y} \! \mid \! B_{i})^{s}}{ \sum_{\boldsymbol{b}\in{\mathbb{B}^m}} \! \Pi_{i=1}^m P_{B_i} (b_{i})  q_{\boldsymbol{Y} \mid B_i} (\boldsymbol{Y} \! \mid \! b_{i})^{s} } \right] \\
	\,\,\,\, = \mathbb{E}_{\boldsymbol{B},\boldsymbol{Y}} \! \left[  \log_2 \Pi_{i=1}^m P_{B_i}(B_{i}) \right] - \mathbb{E}_{\boldsymbol{B},\boldsymbol{Y}} \! \left[ \log_2 P_{\boldsymbol{B}}(\boldsymbol{B}) \right] \nonumber \\
	\,\,\,\,\,\,\,\,\,\, + \mathbb{E}_{\boldsymbol{B},\boldsymbol{Y}} \!\! \left[  \log_2 \frac{\Pi_{i=1}^m q_{\boldsymbol{Y} \mid B_i} (\boldsymbol{Y} \! \mid \! B_{i})^{s}}{ \sum_{\boldsymbol{b}\in{\mathbb{B}^m}} \! \Pi_{i=1}^m P_{B_i} (b_{i})  q_{\boldsymbol{Y} \mid B_i} (\boldsymbol{Y} \! \mid \! b_{i})^{s} } \right] \\
	\,\,\,\, = \sum_{i=1}^m \mathbb{E}_{\boldsymbol{B}} \left[ \log_2 P_{B_i}(B_{i}) \right] - \mathbb{E}_{\boldsymbol{B}} \! \left[ \log_2 P_{\boldsymbol{B}}(\boldsymbol{B}) \right]  \nonumber \\
	\,\,\,\,\,\,\,\,\,\, + \mathbb{E}_{\boldsymbol{B},\boldsymbol{Y}} \!\! \left[  \log_2 \frac{\Pi_{i=1}^m q_{\boldsymbol{Y} \mid B_i} (\boldsymbol{Y} \! \mid \! B_{i})^{s}}{ \Pi_{i=1}^m \sum_{b \in{\mathbb{B}}} P_{B_i} (b)  q_{\boldsymbol{Y} \mid B_i} (\boldsymbol{Y} \! \mid \! b)^{s} } \right] \\
	\label{eq:Iqgs1}
	\,\,\,\, = - \sum_{i=1}^m \mathbb{H}(B_i) + \mathbb{H}(\boldsymbol{B}) + \sum_{i=1}^m I_{q_i,s}^{\text{gmi}}(B_i; \boldsymbol{Y}),
\end{IEEEeqnarray}
where the third term is shown in \cite[Eq.~(4.47)]{bicmbook}, i.e., 
\begin{IEEEeqnarray}{l}
\label{eq:Iqgs_rfm}
	\!\!\!\!\!\!\!\! I_{q_i,s}^{\text{gmi}} (B_i; \boldsymbol{Y}) \! = \! \mathbb{E}_{B_i,\boldsymbol{Y}} \!\! \left[  \log_2 \frac{q_{\boldsymbol{Y} \mid B_i} (\boldsymbol{Y} \! \mid \! B_{i}) ^s}{\sum_{b \in{\mathbb{B}}} P_{B_i} (b)  q_{\boldsymbol{Y} \mid B_i} (\boldsymbol{Y} \! \mid \! b) ^s} \right] .
\end{IEEEeqnarray}
By employing \cite[Th.~4.20]{bicmbook}, \eqref{eq:Iqgs_rfm} is reformulated to
\begin{IEEEeqnarray}{L}
\label{eq:Iqgs1b}
	\!\!\!\!\!\!\!\! I_{q_i,s}^{\text{gmi}} (B_i; \boldsymbol{Y}) = \mathbb{H}(B_i) - \sum_{b\in\mathbb{B}} P_{B_i}(b) \cdot \nonumber \\
	\!\! \mathbb{E}_{L_i^{\text{ex}} \mid B_i=b} \! \left[ \log_2 (1+\exp (-(-1)^b \cdot ( s \hat{L}_i^{\text{ex}}(\boldsymbol{Y}) + L_i^{\text{pr}} ))) \right] 
\end{IEEEeqnarray}
Applying \eqref{eq:Iqgs1b} and \eqref{eq:Lvalue} to \eqref{eq:Iqgs1}, we obtain
\begin{IEEEeqnarray}{L}
\label{eq:Iqgs1e}
	I_{q,s}^{\text{gmi}} (\boldsymbol{B}; \boldsymbol{Y}) = \mathbb{H}(\boldsymbol{B}) -  \sum_{i=1}^m \sum_{b\in \mathbb{B}} P_{B_i}(b) \cdot \nonumber \\
	\,\,\,\,\,\,\,\, \mathbb{E}_{L_i^{\text{po}} \mid B_i = b} \! \left[ \log_2 (1+\exp (-(-1)^b \! \cdot \!  \hat{L}_i^{\text{po}}(\boldsymbol{Y}) )) \right] .
\end{IEEEeqnarray}
According to \eqref{eq:asi_mc_2}, \eqref{eq:Iqgs1e}, and $s = s_{\text{o}}$, the ASI equals to the NGMI in \eqref{eq:NGMIdef}.

\section*{Appendix C \\ ASI from quantized L-values}
For quantized L-values $l \in \mathbb{L}$, ASI is computed by
\begin{IEEEeqnarray}{rCL}
	\label{eq:ASI_quant}
	\text{ASI} & = & 1 - \mathbb{H}(L_{\text{a}} \! \mid \! |L_{\text{a}}|) = 1 + \mathbb{H}(|L_\text{a}|) - \mathbb{H}(L_{\text{a}}) \\
	\label{eq:ASI_quant2}
	& = & 1 - \sum_{l \in \mathbb{L}} P_{L_{\text{a}}}(l) \log_2 \left(1+\frac{P_{L_{\text{a}}}(-l)}{P_{L_{\text{a}}}(l)} \right) 
\end{IEEEeqnarray}
based on the pmf $P_{L_{\text{a}}}(l)$. Considering \eqref{eq:La_ratio}, for $l \in \mathbb{L} \setminus \{ \pm l_{\text{max}} \}$,
\begin{IEEEeqnarray}{rCL}
	\frac{P_{L_{\text{a}}}(-l)}{P_{L_{\text{a}}}(l)} & = & \frac{\int_{l'=l-\frac{\Delta l}{2}}^{l+\frac{\Delta l}{2}}p_{L_{\text{a}}} (l') \text{e}^{-\frac{s_{\text{o}}}{s} l'} \text{d}l'}{\int_{l'=l-\frac{\Delta l}{2}}^{l+\frac{\Delta l}{2}}q_{L_{\text{a}}} (l') \text{d}l'} \\
	& \approx & \frac{\Delta l \cdot p_{L_{\text{a}}}(l) \text{e}^{-\frac{s_{\text{o}}}{s} l} \cdot \frac{ \text{e}^{\frac{\Delta l}{2}} + \text{e}^{-\frac{\Delta l}{2}} }{2} }{\Delta l \cdot p_{L_{\text{a}}}(l)} \\
	\label{eq:QLa_ratio}
	& = & \text{e}^{-\frac{s_{\text{o}}}{s} l} \cdot \frac{ \text{e}^{\frac{\Delta l}{2}} + \text{e}^{-\frac{\Delta l}{2}} }{2} . 
\end{IEEEeqnarray}
For $l \in \{ \pm l_{\text{max}} \}$,
\begin{IEEEeqnarray}{rCL}
	\frac{P_{L_{\text{a}}}(-l)}{P_{L_{\text{a}}}(l)} & = & \frac{\int_{l'=l-\frac{\Delta l}{2}}^{\infty}p_{L_{\text{a}}} (l') \text{e}^{-\frac{s_{\text{o}}}{s} l'} \text{d}l'}{\int_{l'=l-\frac{\Delta l}{2}}^{\infty}p_{L_{\text{a}}} (l') \text{d}l'} .
\end{IEEEeqnarray}
At a reasonable choice of the resolution $n_{\text{L}}$ and the quantization step $\Delta l$, $P_{L_{\text{a}}}(l_{\text{max}}) \gg P_{L_{\text{a}}}(-l_{\text{max}}) \approx 0$. In such cases, $P_{L_{\text{a}}}(l)\,(l\in \{ \pm (l_{\text{max}}) \})$ does not influence \sout{to} the ASI in \eqref{eq:ASI_quant2}. Then, applying \eqref{eq:QLa_ratio} to \eqref{eq:ASI_quant2}, the ASI is approximately computed via Monte-Calro integration, i.e., 
\begin{IEEEeqnarray}{L}
	\label{eq:ASI_quant_mc}
	\!\!\!\!\!\! \text{ASI} \approx 1 - \mathbb{E}_{L_{\text{a}}} \! \left[ \log_2 \! \left( 1 + \text{e}^{- \frac{s_{\text{o}}}{s} L_{\text{a}} } \cdot \frac{ \text{e}^{\frac{\Delta l}{2}} + \text{e}^{-\frac{\Delta l}{2}} }{2}  \right) \right] .
\end{IEEEeqnarray}

\balance


\end{document}